\documentclass[10pt,conference,comsocconf,letterpaper]{IEEEtran}
\makeatletter
\def\ps@headings{%
\def\@oddhead{\mbox{}\scriptsize\rightmark \hfil \thepage}%
\def\@evenhead{\scriptsize\thepage \hfil \leftmark\mbox{}}%
\def\@oddfoot{}%
\def\@evenfoot{}}
\makeatother
\usepackage{amsmath,amssymb}
\usepackage{comment}
\usepackage{amsmath,bm}
\usepackage[]{algorithm, algorithmic}
\usepackage{graphicx,subfigure}
\usepackage{textcomp}
\usepackage{graphicx}

\begin{document}

\title{Cost Minimization in Multiple IaaS Clouds: A Double Auction Approach}\vspace{-15mm}
\author{\IEEEauthorblockN{
	Jian Zhao\IEEEauthorrefmark{1},
	Chuan Wu\IEEEauthorrefmark{1}, Zongpeng Li\IEEEauthorrefmark{2}}\\\vspace{-3.5mm}
	\IEEEauthorblockA{\IEEEauthorrefmark{1} The University of Hong
	Kong, Hong Kong, 
	\{jzhao,cwu\}@cs.hku.hk}
	\IEEEauthorblockA{\IEEEauthorrefmark{2} University of Calgary, 
	zongpeng@ucalgary.ca}\vspace{-10mm}
}

\maketitle

\begin{abstract}
IaaS clouds invest substantial capital in operating their data centers. Reducing the cost of resource provisioning, is their forever pursuing goal. Computing resource trading among multiple IaaS clouds provide a potential for IaaS clouds to utilize cheaper resources to fulfill their jobs, by exploiting the diversities of different clouds' workloads and operational costs. In this paper, we focus on studying the IaaS clouds' cost reduction through computing resource trading among multiple IaaS clouds. We formulate the global cost minimization problem among multiple IaaS clouds under cooperative scenario where each individual cloud's workload and cost information is known. Taking into consideration jobs with disparate lengths, a non-preemptive approximation algorithm for leftover job migration and new job scheduling is designed. Given to the selfishness of individual clouds, we further design a randomized double auction mechanism to elicit clouds' truthful bidding for buying or selling virtual machines. We evaluate our algorithms using trace-driven simulations.

\end{abstract}

\section{Introduction} \label{sec:introduction}
Cloud computing is emerging as a new paradigm that offers users on-demand access to computing resources with small management overhead. As a basic model for delivering cloud services, Infrastructure-as-a-Service (IaaS) has been adopted for serving the ever-growing demand of computing, as exemplified by Amazon EC2 \cite{AMAZONEC2}, Windows Azure \cite{WindowsAzure}, Google Compute Engine \cite{GCE}, Datapipe \cite{Datapipe} and HP Cloud \cite{HPCLOUD}. Such IaaS clouds run data centers with a ``sea'' of computing servers, and face a significant challenge of minimizing the operational cost incurred, while satisfying job requests from users \cite{CutElectricBill_Sigcomm09} \cite{Greenberg2009}. Extensive research has been devoted to reducing the power cost in data centers \cite{CutElectricBill_Sigcomm09,2Tscale_Yao2012}, often with a focus on algorithm design that helps a single cloud provider with cost minimization, through strategies such as CPU speed scaling, dynamic server capacity provisioning, or exploiting the temporal and geographical diversities in power cost in a single cloud's geo-distributed data centers.


Such diversities are actually more evident and pronounced across different IaaS clouds, given the typically limited geographic span of a single cloud. Both workload and resource cost of a cloud vary with time, and such variation curves often complement each other across different clouds distributed in distinct geographical areas, manifesting opportunities for inter-cloud job scheduling and migration that enable significant savings in operational cost. However, such opportunities cannot materialize without the help of (i) a judicious scheduling algorithm that computes the optimal scheduling decisions, and (ii) an effective market mechanism that elicits desirable behaviors from individual clouds with their own economic interests, for such decisions to be implemented.

Users of IaaS clouds are actually already exploiting such diversities among different IaaS clouds to reduce their cost. For example, the cloud management platform RightScale \cite{RightScale} enables its users to efficiently, automatically provide and operate their resources across different IaaS clouds through a configuration framework, {\em the ServerTemplate}. Showing the practical feasibility of deploying computing jobs and resources across IaaS clouds, such realworld applications indeed preceded and motivated our theoretical investigation in this work.

We first model a global time-averaged cost minimization problem for a federation of multiple IaaS clouds under the cooperative scenario, where the solution space includes strategies for scheduling newly arriving jobs and migrating existing jobs. A one-shot optimization problem is derived based on the Lyapunov optimization framework, with an individual cloud's job queue backlogs representing its workload information, and the virtual machine (VM) and network cost for completing jobs as the operational cost. The one-shot optimization is a non-linear integer programming problem in general. However, practical solutions often prefer to place all VM instances of a single job in the same cloud, leading to a linear integer program instead. We design an approximation algorithm for the one-shot optimization for this practical scenario, guaranteeing a solution that is within a constant gap $C$ from the optimum to the original non-linear integer problem. 
 By running this approximation algorithm in each time slot, we can guarantee that the time-averaged system-wide cost approaches the long-term offline optimum asymptotically. 

Real-world clouds are autonomous entities with their own economic interests, and the workload information in terms of job queue backlogs and realtime operational cost are private. A truthful market mechanism is needed, for eliciting truthful reports on such information from them, {\em e.g.}, through buy-bids and sell-bids for VM instances in an auction. We design a randomized double auction mechanism that is truthful in expectation. The winner determination problem (WDP) of the double auction under truthfully bidding is equivalent to the one-shot optimization problem in the cooperative scenario. Our cooperative approximation algorithm can compute a solution to the WDP that both (a) guarantees an approximation ratio
of $\frac{1}{1-\delta}$, and (b) verifies the integrality gap between the WDP integer program (IP) and its linear programming relaxation (LPR) with the same upper-bound $\frac{1}{1-\delta}$.

Our double auction mechanism operates in each time slot as follows. First, we simulate a fractional Vickrey-Clarke-Grove (VCG) auction with the LPR of the WDP as the social welfare maximization problem, and compute an optimal fractional VM allocation together with fractional VCG charges and payments. Second, we apply a primal-dual decomposition technique for packing-type optimization problems due to Carr {\em et al.} \cite{Carr_RandomizedMetarounding} and Lavi {\em et al.} \cite{Lavi_RandomizedAuction} for decomposing the fractional optimal allocation, scaled down by $1-\delta$, into a weighted combination of integral allocations. Third, we randomly pick each integral solution in the combination with probability equaling its weight. Finally, we scale down the fractional VCG charges and payments by $1-\delta$ to be the charges and payments expectation of our randomized double auction. We show that the resulting auction is not only truthful in expectation, but simultaneously achieves the same approximation ratio $\frac{1}{1-\delta}$ to the WDP, and therefore can be applied as a cloud market mechanism for solving the one-shot optimization for achieving time-averaged long-term minimum system cost.

The fact that our randomized auction essentially guarantees the same approximation ratio to the WDP as the cooperative approximation algorithm does is somewhat notable. The cooperative algorithm can assume truthful bids for free and focus on algorithmically optimizing the WDP objective, while our auction has to pay close attention to truthfulness in both sell-bids and buy-bids. Indeed, most existing truthful auctions that are based on approximate social welfare maximization result in an extra factor loss in social welfare. To our knowledge, this work represents the first double auction that is truthful in both sides, and matches the cooperative social welfare approximation ratio without loss.

In the rest of the paper, Sec.~\ref{sec:relatedwork} reviews related literature. Sec.~\ref{sec:model} presents system model. Sec.~\ref{sec:lyapunov}  derives the one-shot minimization problem. Sec.~\ref{sec:method} designs an approximation algorithm for the one-shot optimization. Sec.~\ref{sec:game} presents the randomized double auction. Sec.~\ref{sec:simulations} presents our simulation results, and Sec.~\ref{sec:conclusion} concludes the paper.

\section{Related Work} \label{sec:relatedwork}


The cost minimization problem has been a focus of studies in the cloud computing literature due to its economic importance \cite{2Tscale_Yao2012}\cite{CutElectricBill_Sigcomm09}\cite{MinElectricCost_Infocom10}\cite{GreenLoadBalance_Sigmetrics11}. These studies often focus on a single cloud with geographically distributed data centers. Qureshi {\em et al.} \cite{CutElectricBill_Sigcomm09} develop an electricity price aware workload routing method for reducing the electric bill. Rao {\em et al.} \cite{MinElectricCost_Infocom10} formulate an electricity cost minimization problem as a constrained mixed-integer programming problem under the temporal and spatial diversities of electricity prices. Liu {\em et al.} \cite{GreenLoadBalance_Sigmetrics11} take both delay cost and electricity cost into consideration for workload routing and processing in a cloud's distributed data centers. Yao {\em et al.} \cite{2Tscale_Yao2012} use a two-time-scale framework for integrating decisions on CPU speed scaling, server right-sizing and inter-data center workload routing among a cloud's distributed data centers, for cost reduction. This work instead pursues provable long-term cost reduction in a cloud federation through a randomized double auction mechanism.


For resource trading among multiple clouds, Mihailescu {\em et al.} \cite{DynPricingFederated_CCGrid10,RationalityImpactFederated_CCGrid12} discuss the advantages of dynamic pricing in resource trading among a cloud federation. Although the economical and computational advantages of dynamic pricing are illustrated, they are not enabled by a practical market mechanism such as a truthful auction. Li {\em et al.} \cite{VMTrading_Infocom13} study the resource trading among multiple IaaS clouds and design a double auction mechanism. Different from this work, they focus on the revenue an individual cloud can glean from the auction, instead of system-wide cost savings. 




Lavi {\em et al.} \cite{Lavi_RandomizedAuction} propose a single-sided randomized auction developed upon a primal-dual LP decomposition technique due to Carr {\em et.al.} \cite{Carr_RandomizedMetarounding}, for combinatorial optimization problems with an underlying packing structure. Such single-sided randomized auctions have been applied in the secondary spectrum market \cite{SecondaryNetworksAuction_Infocom12} \cite{MultihopSecondaryNetworksAuction_TMC}, for efficient channel allocation. To the authors' knowledge, this work is the first that generalizes such a randomized, truthful auction framework into the double auction paradigm.

\section{Model and Notation} \label{sec:model}

\vspace{-1mm}
\subsection{System Model}

\vspace{-1mm}
We consider a number $J$ of federated IaaS clouds that run in a time slotted manner. We assume that the IaaS clouds support interoperability through standardizations such as the Open Virtualization Format (OVF) \cite{OVF}, Open Cloud Computing Interface (OCCI) \cite{OCCI}, such that jobs can be readily migrated across different IaaS clouds \cite{interoperabilitysurvey}. Job scheduling and migration decisions are made and executed at the beginning of each time slot $t \in [0,T]$.
We assume that each cloud provider $j \in [1,J]$ operates a single data center for ease of presentation; our techniques and results can be generalized to clouds with multiple data centers in a straightforward way.


Each cloud $j$ has $N_j$ homogeneous servers that each can provide $H_j$ VM instances. The cost of providing one VM instance at time slot $t$ in cloud $j$ is $\beta_j(t)$.
Clouds are connected through network links leased from Internet Service Providers (ISPs).
The cost for transferring a unit data out of cloud $i$ to cloud $j$ is $\lambda_i$, which is dependent on the source cloud only, as is the case in real-world clouds such as Amazon EC2 \cite{AMAZONDTpricing} 
and Windows Azure \cite{WindowsAzureDTpricing}. 

Each IaaS cloud receives jobs from its customers. A job is specified by a pair $(g^k, w^k)$, where $g^k$ is the number of VMs requested, and $w^k \in [w^{min}, w^{max}]$ is the number of time slots that these VMs are needed for. Let $K$ be the total number of job types. We define the workload of a job as $g^k\cdot w^k$. Each cloud $j$ maintains a queue for unscheduled jobs of type $k$, with queue backlog $q_j^k(t)$, for each $k\in [1,K]$. 

\vspace{-1mm}
\subsection{Job Migration \& New Job Scheduling}
In each time slot, a cloud faces two types of decisions: (1) for each newly arrived job, whether it should be admitted and which cloud's resources should be exploited to execute the job; 
(2) for each leftover job from the previous time slot, whether it should be migrated to another cloud for continuing execution or not.

Let $\widetilde{\mathcal{U}}_j(t)$ denote the set of leftover jobs of cloud provider $j$ in $t$, which may have been executed at different clouds, and $\widetilde{U}_j(t) = |\widetilde{\mathcal{U}}_j(t)|$.
Let $\mathcal{U}_j^k(t)$ denote the set of type-$k$ jobs of cloud provider $j$ that are newly scheduled to start execution in $t$ (and can be scheduled to run on different clouds), and $U_j^k(t) = |\mathcal{U}_j^k(t)|$. The actual number of newly scheduled type-$k$ jobs is $\widetilde{U}_j^k(t) = \min \{ q_j^k(t), U_j^k(t) \}$. In the following, we use $\widetilde{\mathcal{U}}_j$, $\mathcal{U}_j^k$ instead of $\widetilde{\mathcal{U}}_j(t)$, $\mathcal{U}_j^k(t)$ when there is no confusion.

\noindent\textbf{VM Placement. } A cloud $j$ decides on the leftover job migration and new job scheduling by computing the placement of VMs belonging to a job $l\in \cup_{\mathcal{K}}\mathcal{U}_j^k \cup \widetilde{\mathcal{U}}_j$. Let the VMs in a type-$k$ job be numbered from $1$ to $g^k$. Each job $l$ of type $k$ is associated with a $g^k \times g^k$ traffic matrix $T^l(t)$, with the entry $T^l_{r,s}(t)$ being the traffic volume from VM $r$ to VM $s$.  

We use a $J\times 1$ vector $I^l_s(t)$ to denote the placement of job $l$'s VM $s, s\in [1,K]$, at time slot $t$, and its $i$-th entry is

\vspace{-3mm}
{\small\begin{align}\label{con:indicator}
I^l_{i,s}(t) = \left\{\begin{array} {ll}
1 & \mbox{ if job $l$'s VM $s$ is placed in cloud $i$ at time $t$; } \\
0 & \mbox{ otherwise.}
\end{array}\right.
\end{align}
}
\vspace{-4mm}

A $J\times g^k$ matrix $I^l = [I^l_1,\ldots, I^l_{g^k} ]$ indicates the placement of all VMs of job $l$. We use $I^l(t-1), l\in \widetilde{\mathcal{U}}_j$ to denote the locations of leftover jobs of cloud $j$ at the beginning of time slot $t$. We need to determine the placement of all VMs of a job $l\in \cup_{\mathcal{K}}\mathcal{U}_j^k \cup \widetilde{\mathcal{U}}_j$ at time slot $t$, {\em i.e.}, $I^l(t)$.

Each VM $s$ of job $l$ is placed in exactly one cloud

\vspace{-4mm}
{\small\begin{align}
&\sum_{i=1}^J I^l_{i,s}(t) = 1, \quad \forall l, \forall s 
\label{con:placement2}
\end{align}
}
\vspace{-4mm}

Furthermore, VMs placed in the same cloud should not exceed the capacity of the latter,

\vspace{-4mm}
{\small\begin{align}
\sum_{j\in\mathcal{J}}\sum_{l\in \widetilde{\mathcal{U}}_j }\sum_{s=1}^{g^k} I^l_{i,s}(t) + \sum_{j\in\mathcal{J}}\sum_{l\in \cup_{\mathcal{K}}\mathcal{U}_j^k }\sum_{s=1}^{g^k} I^l_{i,s}(t) \le N_iH_i, i\in [1,J]
\label{placement}
\end{align}
}
\vspace{-4mm}

\noindent \textbf{Job Queue Updating. }
The federated clouds provide a common Service Level Objective (SLO) guarantee on service response time, from when the job arrives to when it starts to run on VMs. In practice, SLOs can be any specific measurable characteristics of the Service Level Agreement (SLA), such as availability, throughput, frequency, response time, or quality. Here we focus on studying the service response time as the SLO. This common SLO can be expressed in the standard way with the Open Cloud Computing Interface \cite{interoperabilitysurvey}: 

\vspace{-4mm}
{\small\begin{align}
&\mbox{Each job is either served or dropped (subject to a penalty)} \nonumber \\
&\mbox{within the maximum response delay $d$.} \label{con:delay}
\end{align}
}
\vspace{-4mm}

Upon violation of the SLO, the cloud provider pays a penalty $\alpha_j^k$ and drops the job.\footnote{In practice, a cloud may never drop a user's job. The ``drop'' in our model can be understood as follows: Each cloud $j$ maintains a set of regular resources ($N_iH_i$ VMs) while keeping a set of backup resources, whose provisioning can be expensive. When a job is ``dropped'' due to not being scheduled using the regular resources when its response delay is due, the cloud uses its expensive backup resources to serve the job, subject to a cost $\alpha_j^k$ (``the job drop penalty'') to serve one type-$k$ job.}
Let $A_j^k(t)\in[0,A^{max}]$ ($A^{max}$ is an upper bound for one type of job arrivals in each cloud) and $G_j^k(t)$ be the number of type-$k$ jobs cloud $j$ receives from its customers and drops at time $t$, respectively. The updates of the job queues in cloud $j$ are:

\vspace{-4mm}
{\small\begin{align}
q_j^k(t+1) = \max\{ q_j^k(t)-U_j^k(t)-G_j^k(t), 0 \} + A_j^k(t), k\in[1,K].
\label{con:dynamicQqueue}
\end{align}
}\vspace{-4mm}

\noindent Here {\small$U_j^k(t) = \sum_{l\in \mathcal{U}_j^k}\sum_{s=1}^{g^k} \sum_{i=1}^J I^l_{i,s}(t)/g^k$} and

\vspace{-4mm}
{\small\begin{align}
0 \le G_j^k(t) \le G_j^{k,max}, k\in [1,K], \label{con:max_drop}
\end{align}
}
\vspace{-1mm}
\noindent where $G_j^{k,max}$ is an upper bound of $G_j^k(t)$.
We apply the $\epsilon$-persistent service queue technique \cite{Lya-2} to guarantee constraint (\ref{con:delay}). We associate each job queue $q_j^k$ with a virtual queue $Z_j^k$, with initial backlog $Z_j^k(0) = 0$, updated by

\vspace{-4mm}
{\small\begin{align}
Z_j^k(t+1) =  & \max \left\{ Z_j^k(t) + 1_{q_j^k(t)>0} \cdot [\epsilon_j^k - U_j^k(t)] \right.\nonumber \\
& \left.- G_j^k(t) - 1_{q_j^k(t)=0}\cdot U^{k,max}_j, 0 \right\}, j\in[1,J], k\in [1,K].
\label{con:dynamicZqueue}
\end{align}
}
\vspace{-4mm}

Here the indicator function $1_{q_j^k(t)>0}$  is 1 when $q_j^k(t) > 0$, and $0$ otherwise. Similar for $1_{q_j^k(t)=0}$. The virtual queue is built in such a way to make sure its departure rate equals to that of the corresponding job queue when the queue backlog of job queue is larger than $0$. A predefined constant $\epsilon_j^k \le A^{max}$ can be gauged to control the queueing delay bound. Any algorithm that maintains bounded $Z_j^k(t)$ and $q_j^k(t)$  ensures jobs in queue $q_j^k(t)$ are served within bounded worst-case delay. The rationale can be explained as follows. Consider jobs arriving at $t$. If $q_j^k$ reaches $0$ within the subsequent $d$ time slots, the jobs are served within $d$ time slots. Otherwise, $Z_j^k$ has a constant arrival rate $\epsilon_j^k$, and the same departure rate with $q_j^k$, {\em i.e.}, $U_j^k(t) + G_j^k(t)$. Let $Z_j^{k,max}$ and $q_j^{k,max}$ be the upper bound of the size of queue $Z_j^k$ and $q_j^{k}$ at any time, respectively. For the $d$ time slots following $t$, the total arrival into $Z_j^k$ minus the total departure is {\small $\epsilon_j^k d - \sum_{\tau=t+1}^{t+d}[U_j^k(\tau) + G_j^k(\tau)]$}, as {\small $Z_j^k$} is bounded by {\small $Z_j^{k,max}$}, {\small $\epsilon_j^k d - \sum_{\tau=t+1}^{t+d}[U_j^k(\tau) + G_j^k(\tau)] \le Z_j^{k,max}$}. When {\small $d = \lceil (Z_j^{k,max} + q_j^{k,max})/\epsilon_j^k \rceil$}, we have {\small $\epsilon_j^k d - Z_j^{k,max} \ge q_j^{k,max}$}, hence, {\small $\sum_{\tau=t+1}^{t+d}[U_j^k(\tau) + G_j^k(\tau)] \ge q_j^{k,max} \ge q_j^k(t+1)$}. As job queues are FIFO, all jobs arriving at time slot $t$ will be served within a delay of {\small $d = \lceil (Z_j^{k,max} + q_j^{k,max})/\epsilon_j^k \rceil$} time slots.

\vspace{-1mm}
\subsection{Cost Minimization Problem Formulation}

The inter-cloud traffic includes two types: (i) traffic among VMs of the same job placed in different clouds, and (ii) traffic due to VM migration of leftover jobs. The type (i) traffic from cloud $i$ to cloud $j$ due to job {\small $l\in \cup_{\mathcal{K}} \mathcal{U}_j^k \cup \widetilde{\mathcal{U}}_j $}, is {\small $\sum_{r=1}^{g^k}\sum_{s = 1} ^{g^k} I^l_{i,r}(t) \cdot I^l_{j,s}(t)\cdot T^l_{r,s}(t)$}. It is the $i\times j^{\mbox{th}}$ entry of matrix {\small $[I^l(t)\cdot T^l(t)\cdot I^l(t)']$, {\em i.e.}, $[I^l(t)\cdot T^l\cdot I^l(t)']_{i,j}$}, where $I^l(t)'$ is the transpose of $I^l(t)$. When {\small $l\in \widetilde{\mathcal{U}}_j$}, job $l$ is a leftover job. $I^l_s(t-1)$ is the indicator of the placement of job $l$'s VM $s$ at time slot $t-1$. VM $s$ of job $l$ is migrated if {\small $I^l_s(t)\cdot I^l_s(t-1) = 0$}, and is not migrated otherwise.

For type (ii) traffic, we assume that only the up-to-date virtual machine states need to be transferred for VM migration, such that the temporary VM downtime during migration is short. 
Let $\phi^k$ be the data volume involved when migrating one VM of a type-$k$ job. The migration traffic from cloud $i$ to cloud $j$ due to job $l$'s VMs is {\small $\sum_{s=1}^{g^k}I^l_{i,s}(t-1) \cdot I^l_{j,s}(t)\cdot \phi^k$}. It is the {\small  $i\times j^{\mbox{th}}$} entry of matrix {\small $[I^l(t-1)\cdot I^l(t)'\cdot \phi^k]$}, {\em i.e.}, {\small $[I^l(t-1)\cdot I^l(t)'\cdot \phi^k]_{i,j}$}.

At time slot $t$, the cost for running a leftover job $l\in\widetilde{\mathcal{U}}_j$, for running a newly scheduled job $l\in\mathcal{U}_j^k$, and for dropping a job $l$ is $c^l =$

\vspace{-3mm}
{\small\begin{align*}
\left\{\begin{array} {l}
\sum_{i=1}^J\beta_i(t)\sum_{s=1}^{g^k}I^l_{i,s}(t) + \sum_{h\neq i}\sum_{i=1}^J \lambda_{h} [I^l(t)\cdot T^l\cdot I^l(t)']_{h,i}  \\
+ \sum_{h\neq i}\sum_{i=1}^J \lambda_{h}\sum_{s=1}^{g^k}I^l_{h,s}(t-1) \cdot I^l_{i,s}(t)\cdot \phi^k, \\
\quad\quad\quad\quad\quad\quad \mbox{if } l\in\widetilde{\mathcal{U}}_j, j\in[1,J]; \\
\sum_{i=1}^J\beta_i(t)\sum_{s=1}^{g^k}I^l_{i,s}(t) + \sum_{h\neq i}\sum_{i=1}^J\lambda_h[I^l(t)\cdot T^l\cdot I^l(t)']_{h,i},  \\
\quad\quad\quad\quad\quad\quad \mbox{if } l\in\mathcal{U}_j^k, j\in[1,J], k\in[1,K]; \\
\alpha_j^k, \quad\quad\quad\quad\quad \mbox{if $l$ is dropped.}
\end{array}\right.
\end{align*}
}
\vspace{-2mm}

The time-averaged cost incurred by all cloud $j$'s jobs is,

\vspace{-4mm}
{\footnotesize\begin{align*}
C_j(t) = \lim_{T\to \infty} \frac{1}{T}\sum_{t=0}^{T-1}\mathbb{E} \left\{\sum_{l\in \widetilde{\mathcal{U}}_j}c^l + \sum_{l\in \cup_{\mathcal{K}}\mathcal{U}_j^k}c^l+ \sum_{k=1}^K\alpha^k_j G^k_j(t) \right\}
\end{align*}
}
\vspace{-2mm}



Finally the global cost minimization problem is:

\vspace{-3mm}
{\small\begin{align*}
\min \quad& \sum_{j=1}^J C_j(t) \\
\mbox{s.t. } \quad& (\ref{con:indicator}) - (\ref{con:delay}).
\end{align*}
}
\vspace{-4mm}

The decision variables are $U_j^k(t)$, $I^l_{i,s}(t)$, and $G_j^k(t)$, $\forall i,j\in [1,J], k\in [1,K], l\in \cup_{\mathcal{K}}\mathcal{U}_j^k \cup \widetilde{\mathcal{U}}_j, s\in [1,g^k], t\in [0,T]$. 
Notations are summarized in a notation table below for ease of reference.

\begin{table}[h]
\vspace{-2mm}
\begin{tabular}{|l|p{3.19cm}|l|p{3.18cm}|}
\hline
$K$ & \# of job types &  $N_j$ & \# of servers in cloud $j$ \\ \hline
$w^k$ & service time of type-$k$ job & $Z_j^k$ & virtual queue length \\ \hline
$d$ & max response delay & $T^l$ & traffic matrix of job $l$ \\ \hline
$\Gamma$ & a time frame & $c^l$ & cost of job $l$ \\ 
\end{tabular}
\begin{tabular}{|c|p{7cm}|}
	\hline
$J$ & total \# of IaaS cloud providers in the federation.  \\ \hline
$H_j$ & \# of instances a server in cloud $j$ can host.  \\ \hline
$\beta_j(t)$ & cost of hosting a VM in cloud $j$. \\ \hline
$\lambda_i$ & cost of transferring a unit volume of data out of cloud $i$. \\ \hline
$g^k$ & \# of VMs required by type-$k$ job. \\ \hline
$A_j^k(t)$ & \# of type-$k$ jobs arriving at $t$ in cloud $j$. \\ \hline
$A^{max}$ & max \# of one type of jobs arriving in one cloud at one time. \\ \hline
$q_j^k$ & queue backlog for unscheduled type-$k$ jobs in cloud $j$. \\ \hline
$U_j^k(t)$ & \# of newly served type-$k$ jobs at $t$ in cloud $j$. \\ \hline
$G_j^k(t)$ & \# of type-$k$ jobs dropped at $t$ in cloud $j$. \\ \hline
$G_j^{k,max}$ & max \# of type-$k$ jobs being dropped at $t$ in cloud $j$. \\ \hline
$I^l_{i,s}$ & 1/0: whether instance $s$ of job $l$ is placed in cloud $i$. \\ \hline
$\widetilde{\mathcal{U}}_j$ & set of leftover jobs from cloud $j$. \\ \hline
$\mathcal{U}_j^k$ & set of newly scheduled type-$k$ jobs from cloud $j$. \\ \hline
$\phi^k$ & data size for migrating one VM of type-$k$ job. \\ \hline
$\mathcal{D}_H$ & set of data centers not allocating VMs to cloud $j^*$. \\ \hline
$\mathcal{D}_L$ & set of data centers allocating VMs to cloud $j^*$. \\ \hline
$\alpha_j^k$ & penalty for dropping one type-$k$ job in cloud $j$. \\ \hline
$\epsilon_j^k$ & parameter used as the incoming rate for virtual queue of $q_j^k$. \\ \hline 
\end{tabular}\label{tab:notation}
\vspace{-4mm}
\end{table}

\vspace{-1mm}
\section{The Lyapunov Framework} \label{sec:lyapunov}
We apply the Lyapunov optimization framework to translate the long-term cost optimization in the cloud federation into one-shot minimization problems. We design a centralized online algorithm for achieving close-to-optimal time-averaged cost by solving the one-shot minimization in Sec.~\ref{sec:method}, and design a double auction mechanism for eliciting desirable cloud behaviours for achieving the same close-to-optimal performance in Sec.~\ref{sec:game}.

\vspace{-1mm}
\subsection{Dealing with Jobs with Varying Lengths}
Departing from existing literature on Lyapunov optimization, we consider jobs with varying lengths, resulting in leftover jobs carried over into a subsequent time slot, complicating optimization decision making in the system.

We group $\Gamma$ time slots into a {\em refresh frame}, where $\Gamma > w^{max}$ and $w^{max}$ is the largest time duration of a job. The time slots can be divided into $R$ consecutive frames. In each frame, only new jobs that can be completed within the frame are taken into consideration for scheduling. Hence, there are no leftover jobs running in data centers at the beginning of each frame. Control decisions from the previous frame are therefore isolated from those in the current frame.

\vspace{-1mm}
\subsection{One-Shot Drift-Plus-Penalty Minimization}
Let $\mathbf{\Theta}_j(t)= (\mathbf{q}_j,\mathbf{Z}_j)$ be the vector of job queues and virtual queues in cloud $j$. $\mathbf{\Theta}(t) =(\mathbf{\Theta}_1(t), \mathbf{\Theta}_2(t),...,\mathbf{\Theta}_J(t))$. Define the {\em Lyapunov function} of $\mathbf{\Theta}_j(t)$ as:

\vspace{-4mm}
{\small\begin{align}
L(\mathbf{\Theta}_j(t)) = \frac{1}{2}\sum_{k=1}^K [(w^k)^2q_j^k(t)^2 + Z_j^k(t)^2]
\end{align}
}
\vspace{-4mm}

The Lyapunov optimization framework guarantees that the long-term global cost minimization in the federated cloud can be achieved by minimizing the following one-shot {\em drift-plus-penalty} under constraints (\ref{con:indicator}) (\ref{con:placement2}) (\ref{placement}) (\ref{con:max_drop}). Detailed derivation of the function is provided in Appendix \ref{sec:appendix_a}.

\vspace{-4mm}
{\small\begin{align}
 \varphi_1(t) + \varphi_2(t)\label{one_slot}
\end{align}
}
\vspace{-4mm}
where

\vspace{-4mm}
{\small\begin{align*}
\varphi_1(t) = & \sum_{j=1}^J\sum_{k=1}^K[ V\alpha_j^k - (w^k)^2q_j^k(t)-Z_j^k(t)]G_j^k(t) \\
\varphi_2(t) = & \sum_{l\in \cup_{\mathcal{J},\mathcal{K}}\mathcal{U}_j^k}Vc^l - \sum_{j=1}^J\sum_{k=1}^K [(w^k)^2q_j^k(t) + Z_j^k(t)]U_j^k(t) \\
& + \sum_{l\in \cup_{\mathcal{J}}\widetilde{\mathcal{U}}_j}Vc^l
\end{align*}
\vspace{-3mm}
}

\noindent and $V$ is a non-negative parameter chosen by the algorithm to tune the tradeoff between cost and service response delay.

\section{One-shot Optimization Problem for Global Cost Minimization} \label{sec:method}
We next decompose the minimization of the one-shot drift-plus-penalty in (\ref{one_slot}) into two independent sub-problems, and solve them using a dynamic algorithm for determining job dropping and another for scheduling existing and new jobs, respectively.
VMs of the same job are placed in the same cloud, for practical feasibility. We prove that our dynamic algorithm approaches optimal one-shot drift-plus-penalty within a constant gap.

\noindent \textbf{Job Dropping. } The first sub-problem of minimizing (\ref{one_slot}) requires cloud providers to make optimal decisions on the number of dropped jobs, by solving the following minimization:

\vspace{-4mm}
{\small\begin{align}
\min        \quad& \varphi_1(t) \nonumber\\
\mbox{s.t. }    \quad& (\ref{con:max_drop})
\label{job_dropping}
\end{align}
}
\vspace{-4mm}

The solution of (\ref{job_dropping}) is:

\vspace{-2mm}
{\small\begin{align}\label{GR_star}
G_j^k(t) = \left\{\begin{array} {ll}
G_j^{k,max}, & \mbox{if } (w^k)^2q^k_j(t) + Z^k_j(t) > V\alpha^k_j; \\
0, & \mbox{if } (w^k)^2q^k_j(t) + Z^k_j(t) \le V\alpha^k_j.
\end{array}\right.
\end{align}
}\vspace{-2mm}

When {\small $V\alpha^k_j < (w^k)^2q^k_j(t) + Z^k_j(t)$}, cloud $j$ can not satisfy SLO for all its jobs, and has to drop some jobs. It is desirable in practice for the cloud provider $j$ to conduct admission control on the maximum job arrival rate $A^{max}$ 
, to avoid job drops. We will assume such admission control for now, and discuss its realization at the end of this section.

\vspace{1mm}
\noindent \textbf{Leftover Job Migration \& New Job Scheduling. } The second sub-problem involves decisions on leftover job migration and new job scheduling that affect $\varphi_2(t)$, as follows: 

\vspace{-4mm}
{\small\begin{align}
\min        \quad& \varphi_2(t) \nonumber\\
\mbox{s.t. }    \quad& (\ref{con:indicator}) (\ref{con:placement2}) (\ref{placement})
\label{migration_scheduling}
\end{align}
}
\vspace{-4mm}

(\ref{migration_scheduling}) is a non-linear integer program (NLIP) 
 with inter-cloud traffic for communication among VMs in the same job (due to the non-linear terms in $c^l$). When VMs of the same job reside in the same cloud, (\ref{migration_scheduling}) becomes a linear integer program as follows,


\vspace{-5mm}
{\small\begin{align}
\min        \quad& \widetilde{\varphi_2}(t) \nonumber\\
\mbox{s.t. }    \quad& (\ref{con:indicator}) (\ref{con:placement2}) (\ref{placement})
\label{migration_scheduling_allinone}
\end{align}
}
\vspace{-5mm}

\noindent and can be solved by the following approximation algorithm.

Now that the VM placement of any VM of a job $l$ is the same, {\em i.e.}, {\small $I_1^l = I_2^l = \ldots = I_{g^k}^l$}, we apply vector $I_0^l$ to represent the placement of job $l$. We have {\small $\sum_{s=1}^{g^k}I^l_{i,s} = g^kI^l_{i,0}$}, and 

\vspace{-4mm}
{\small\begin{align*}
\widetilde{\varphi_2}(t) = & \sum_{l\in \cup_{\mathcal{J},\mathcal{K}}\mathcal{U}^k_j}\sum_{i=1}^J[ \beta_i(t) - \frac{(w^k)^2q_j^k(t) + Z_j^k(t)}{Vg^k} ]Vg^k I^l_{i,0}(t) \\
& + \sum_{l\in \cup_{\mathcal{J}}\widetilde{\mathcal{U}}_j} \sum_{i=1}^J [\beta_i(t) + \sum_{h\neq i} \lambda_{h}I^l_{h,0}(t-1) \cdot \phi^{k_l} ]Vg^{k_l} I^l_{i,0}(t)
\end{align*}
}
\vspace{-3mm}

Variables of $I_{i,0}^l$'s are coupled in constraint (\ref{placement}), the VM capacity constraint at each cloud's data center.
Further analysis reveals that to minimize $\widetilde{\varphi_2}(t)$, available VMs should be allocated to the job queue with the largest value of $\frac{(w^k)^2q_j^k(t) + Z_j^k(t)}{Vg^k}$ among all cloud providers and job types. Let this queue be the queue of type $k^*$ jobs in cloud $j^*$. Jobs from this queue will be scheduled to run in clouds with cost $\beta_i(t) < \frac{(w^{k^*})^2q_{j^*}^{k^*}(t) + Z_{j^*}^{k^*}(t)}{Vg^{k^*}}$. Let $\mathcal{D}_L$ and $\mathcal{D}_H$ denote the set of clouds with cost smaller than and no smaller than $\frac{(w^{k^*})^2q_{j^*}^{k^*}(t) + Z_{j^*}^{k^*}(t)}{Vg^{k^*}}$, respectively. Clouds in $\mathcal{D}_L$ will allocate all their VMs not occupied by leftover jobs, {\em i.e.}, $ N_iH_i - \sum_{l\in \cup_{\mathcal{J}}\widetilde{\mathcal{U}}_j}g^{k_l}I^l_{i,0}(t)$, to serve cloud $j^*$'s type-$k^*$ jobs. Then constraint (\ref{placement}) can be transformed to:

\vspace{-3mm}
{\footnotesize\begin{align}
\sum_{j\in\mathcal{J}}\sum_{l\in \widetilde{\mathcal{U}}_j }g^{k_l} I^l_{i,0}(t) \le N_iH_i, 1\le i\le J;  \label{con:leftover_job} \\
\sum_{j\in\mathcal{J}}\sum_{l\in \cup_{\mathcal{K}}\mathcal{U}_j^k } g^k I^l_{i,0}(t) = 0, \forall i \in \mathcal{D}_H;  \label{con:zeronewjob}\\
\sum_{l\in \mathcal{U}_{j^*}^{k^*}} g^{k^*} I^l_{i,0}(t)
\le N_iH_i - \sum_{j\in\mathcal{J}}\sum_{l\in \widetilde{\mathcal{U}}_j} g^{k_l} I^l_{i,0}(t), \forall i\in\mathcal{D}_L. \label{con:schedulenewjob}
\end{align}
}
\vspace{-3mm}

Constraint (\ref{con:leftover_job}) states that the number of VMs needed by migrated leftover jobs can not exceed the VM capacity at each cloud. Constraint (\ref{con:zeronewjob}) ensures no new job is scheduled to clouds in $\mathcal{D}_H$. Servers in $\mathcal{D}_H$ are turned off except those used for serving leftover jobs. Constraint (\ref{con:schedulenewjob}) implies that as many type-$k^*$ jobs of cloud $j^*$ as possible are scheduled to the available VMs in clouds in $\mathcal{D}_L$.


Substituting (\ref{con:zeronewjob}), (\ref{con:schedulenewjob}) into $\widetilde{\varphi_2}(t)$, we obtain the following optimization problem for leftover job migration,

\vspace{-4mm}
{\footnotesize\begin{align}
\min \quad & \sum_{l\in \cup_{\mathcal{J}}\widetilde{\mathcal{U}}_j} \{\sum_{i\in \mathcal{D}_H} g^{k_l}I^l_{i,0}(t)\cdot[\beta_i(t) + \sum_{h\neq i} \lambda_{h}I^l_{h,0}(t-1) \cdot \phi^{k_l} ] + \sum_{i\in\mathcal{D}_L} g^{k_l}  \nonumber \\
&  \cdot I^l_{i,0}(t)[ \frac{(w^{k^*})^2q_{j^*}^{k^*}(t) + Z_{j^*}^{k^*}(t)}{Vg^{k^*}}¡¡+ \sum_{h\neq i} \lambda_{h}I^l_{h,0}(t-1) \cdot \phi^{k_l} ] \}\nonumber\\
\mbox{s.t. } & (\ref{con:leftover_job})
\label{leftover_job}
\end{align}
}

There are four cases of job migration: (1) When a leftover job in cloud $j\in \mathcal{D}_H$ is migrated to a cloud $i \in \mathcal{D}_L$, the objective function of (\ref{leftover_job}) is decremented by $V\cdot g^{k_l}[\beta_j(t)-\lambda_j\phi^{k_l} - \frac{(w^{k^*})^2q_{j^*}^{k^*}(t) + Z_{j^*}^{k^*}(t)}{Vg^{k^*}} ]$; (2) When a leftover job in $j\in \mathcal{D}_H$ is migrated to $i \in \mathcal{D}_H$, the objective function of (\ref{leftover_job}) is decremented by $V\cdot g^{k_l}[\beta_j(t)-\lambda_j\phi^{k_l} - \beta_i(t)]$; (3) When a leftover job in $j\in\mathcal{D}_L$ is migrated to $i\in \mathcal{D}_H$, the objective function of (\ref{leftover_job}) is incremented by $V\cdot g^{k_l}[\beta_i(t)+\lambda_j\phi^{k_l} - \frac{(w^{k^*})^2q_{j^*}^{k^*}(t) + Z_{j^*}^{k^*}(t)}{Vg^{k^*}}]$; (4) When a leftover job in $j\in\mathcal{D}_L$ is migrated to a $i\in \mathcal{D}_L$, the objective function of (\ref{leftover_job}) is incremented by $V\cdot g^{k_l}\lambda_j\phi^{k_l}$.

Since (\ref{leftover_job}) is a minimization problem, no migration of left-over jobs across clouds in $\mathcal{D}_L$ should happen. For migration of leftover jobs across clouds in $\mathcal{D}_H$, the algorithm first sorts clouds in $\mathcal{D}_H$ from the highest-cost one to the lowest-cost one. It then migrates jobs from higher-cost clouds to lower-cost clouds until no further reduction to (\ref{leftover_job}) is possible.

\begin{algorithm}[htbp]
{\small
\caption{Approximation Algorithm to Solve One-shot Cost Minimization Problem (\ref{migration_scheduling_allinone})}
\label{alg:scheduling}

    \textbf{Input}: $q_j^k(t)$; $Z_j^k(t)$; $I^l(t-1), l \in \widetilde{\mathcal{U}}_j$; $\beta_j(t)$; $\lambda_j$; $\phi^k$; $N_j$; $H_j$; $V$; $\epsilon_j^k$; $\Gamma > w^{max}$; ($\forall k \in \mathcal{K}$, $\forall j \in \mathcal{J}$); $Left$ (Storing sequenced leftover jobs according to value $\beta_{h_l}(t)-\lambda_{h_l}\phi^{k_l}$); $DC$ (Datacenters sorted by cost $\beta_i(t)$).

   { \textbf{Output}: $I^l(t)$, $l$$\in$$\mathcal{U}_j^k$$\cup$$\widetilde{\mathcal{U}}_j$; $G_j^k(t)$, $\forall k$$\in$$\mathcal{K}$, $j$$\in$$\mathcal{J}$} \hfill

\begin{algorithmic}[1]
    \STATE // *** \textbf{Leftover Job Migration} *** //
    \STATE $l = 0$, the leftover job with the highest value; $i=0$, the cloud with the lowest cost. //Initialize $l$ and $i$.

    \WHILE {($l \le \sum_{j=1}^J\widetilde{U}_j$ \& $i \le J$ \& $Left(l).value > DC(i).cost$)}
        \IF {$(DC(i).available \ge Left(l).g)$}  
            \STATE Migrate job $l$ to cloud $i$.
            \STATE Update the number of available VMs in $DC(i)$ by minus $g^{k_l}$.
            \STATE $l++$;
        \ELSE \STATE $i++$;
        \ENDIF
    \ENDWHILE

    \STATE // *** \textbf{New Job Scheduling} *** //
        \FOR {Each job type-$k\in\mathcal{K}$ with $w^k\le \Gamma - (t \mbox{ mod } \Gamma)$ and each cloud provider $j\in\mathcal{J}$}
        \STATE Identify the job queue with the largest $\frac{(w^k)^2q_j^k(t) + Z_j^k(t)}{Vg^k}$, let it be type-$k^*$ jobs in cloud $j^*$.
        \ENDFOR

    \FOR {$\mbox{Each cloud } d\in\mathcal{D}$}
        \IF{$(\frac{(w^{k^*})^2q_{j^*}^{k^*}(t) + Z_{j^*}^{k^*}(t)}{Vg^{k^*}} \le \beta_i(t))$}
            \STATE Keep servers running leftover jobs on; turn off all other servers
        \ELSE
            \STATE Keep servers running leftover jobs on, configure all other servers to run 
 VMs to serve type-$k^*$ jobs of cloud provider $j^*$.
        \ENDIF
    \ENDFOR
    \STATE Decide the number of jobs to drop according to Eqn.~(\ref{GR_star})
\end{algorithmic}}
\end{algorithm}

We summarize our complete algorithm to solve the one-shot minimization problem (\ref{one_slot}) in Algorithm \ref{alg:scheduling}.







\noindent \textbf{Theorem 1.[Approximation Ratio]} {\em Algorithm \ref{alg:scheduling} computes a solution $\widetilde{\varphi_2}^{(a)}(t)$ satisfying $\widetilde{\varphi_2}^{(a)}(t) \le \widetilde{\varphi_2}^*(t) + C \le \varphi_2^*(t) +C $. Here $\widetilde{\varphi_2}^*(t)$ is the optimal solution of the LP relaxation of (\ref{migration_scheduling_allinone}). $\varphi_2^*(t)$ is the optimal solution of (\ref{migration_scheduling}). $C = VJg^{max}\cdot \max\{ \frac{\alpha^{max}}{g^{min}} - \beta_{min} ,\beta_{max}+ \lambda_{max}\phi_{max} - \beta_{min} -\lambda_{min}\phi_{min}\}$. $V$ is the algorithm parameter. $g^{max}$ ($g^{min}$) is the maximum (minimum) number of VMs requested by a job. $\beta_{max}$ ($\beta_{min}$) is the maximum (minimum) cost of one VM. $\lambda_{max}\phi_{max}$ ($\lambda_{min}\phi_{min}$) is the maximum (minimum) cost for migrating one VM. $\alpha^{max}$ is the maximum penalty for dropping one job.
}



Placing all VMs of the same job in the same cloud can lead to waste of VMs in a cloud, which are not sufficient to handle a job.
We observe that the optimal solution of the LP relaxation (LPR) to (\ref{migration_scheduling_allinone}) is smaller than that of (\ref{migration_scheduling}), {\em i.e.}, $\widetilde{\varphi_2}^*(t) \le \varphi_2^*(t)$, because the solution of LP relaxation does not waste VMs due to integer solution constraint and not need to consider inter-VM traffic cost. Algorithm \ref{alg:scheduling} schedules as many jobs as possible to a cloud in $\mathcal{D}_L$. Hence, the number of wasted VMs in one data center is bounded by the maximum number of VMs a job requests, $g^{max}$. The total number of wasted VMs in all $J$ data centers is bounded by $Jg^{max}$. The maximum cost for wasting one VM is $V\cdot\max\{ \frac{\alpha^{max}}{g^{min}} - \beta_{min} ,\beta_{max}+ \lambda_{max}\phi_{max} - \beta_{min} -\lambda_{min}\phi_{min}\}$, it equals to the maximum value by which the one-shot drift-plus-penalty will be deduced if the one VM can be used for serving a job
. Compared with the LP relaxation of (\ref{migration_scheduling_allinone}), Algorithm \ref{alg:scheduling} at most wastes $Jg^{max}$ VMs. The solution of Algorithm \ref{alg:scheduling} is within a constant $C$ from the optimal solution of the LPR of (\ref{migration_scheduling_allinone}), {\em i.e.}, {\small $\widetilde{\varphi_2}^{(a)}(t) \le \widetilde{\varphi_2}^*(t)+ C$}, {\small $C = VJg^{max}\cdot \max\{ \frac{\alpha^{max}}{g^{min}} - \beta_{min} ,\beta_{max}+ \lambda_{max}\phi_{max} - \beta_{min} -\lambda_{min}\phi_{min}\}$}. Hence, we derive {\small $\widetilde{\varphi_2}^{(a)}(t) \le \varphi_2^*(t) +C$}.
The details of the proof is in Appendix \ref{sec:appendix_b}.


\noindent \textbf{Theorem 2.[Conditions for No Job Drop]} {\em When the maximum arrival number for any type of job in one time slot, $A^{max}$, and the total number of VM provisioning in clouds, $\sum_{j=1}^J N_jH_j$, satisfy $Jg^{max}w^{max} \cdot \frac{(w^{max})^2A^{max} + \epsilon^{max}}{g^{min}} < \frac{\Gamma-w^{max}}{\Gamma} \cdot \frac{2\sum_{j=1}^J N_jH_j}{(g^{max})^2}$, i.e., the total arrival rate of all the queues is smaller than $\frac{\Gamma-w^{max}}{\Gamma}$ fraction of the service capacity, the value $[(w^k)^2q_j^k(t) + Z_j^k(t)]/g^k$ will be bounded.
}

\noindent {\em Proof sketch:}
We can see value $[(w^k)^2q_j^k(t) + Z_j^k(t)]/g^k$ as the queue backlog of a new-defined queue corresponding to cloud $j$'s type-$k$ job as follows: in each time slot, the input of the queue is no larger than $[(w^k)^2A^{max} + \epsilon^{max}]/g^k$, $\epsilon^{max} = \max\{\epsilon_j^k, \forall j, k \}$; the output of the queue is no smaller than $2U_j^k/g^k$.
Scheduling of new jobs in Algorithm \ref{alg:scheduling} is a variation of the MaxWeight algorithm. In each time slot it schedules the type of jobs $k^*$ in cloud $j^*$ corresponding to the largest {\small $[(w^{k^*})^2q_{j^*}^{k^*}(t) + Z_{j^*}^{k^*}(t)]/g^{k^*}$}, which can be completed within the current time frame. 
This scheduling algorithm is the same with the Myopic MaxWeight algorithm proposed by Maguluri {\em et.al.} \cite{Maguluri2012}. The total arrival rate for all the queues $[(w^k)^2q_j^k(t) + Z_j^k(t)]/g^k$ is upper-bounded by $Jg^{max}w^{max} \cdot \frac{(w^{max})^2A^{max} + \epsilon^{max}}{g^{min}}$, where $Jg^{max}w^{max}$ is the total number of queues. The total service capacity is larger than $\frac{2\sum_{j=1}^JN_jH_j}{(g^{max})^2}$. Hence, according to Maguluri {\em et al.}'s result \cite{Maguluri2012}, when $Jg^{max}w^{max} \cdot \frac{(w^{max})^2A^{max} + \epsilon^{max}}{g^{min}} < \frac{\Gamma-w^{max}}{\Gamma} \cdot \frac{2\sum_{j=1}^J N_jH_j}{(g^{max})^2}$, {\em i.e.}, the total arrival rate of all the queues is smaller than $\frac{\Gamma-w^{max}}{\Gamma}$ fraction of the service capacity, the queue is strongly stable. The queue backlog will be bounded. As the penalty $\alpha_j^k$ for dropping a type-$k$ job satisfies $ \alpha_j^k \ge [(w^k)^2q_j^k(t) + Z_j^k(t)]/V$, there will be no job drop according to (\ref{GR_star}).

\noindent \textbf{Theorem 3.} {\em When the no job drop condition is satisfied, and the length of a time frame satisfies $\Gamma > w^{max}$, with the assumption that the dynamic virtual machine costs, $\beta_j(t)$, $\forall j\in [1,J]$, and the size for transferring one VM, $\phi^k$, $k\in [1,K]$, are ergodic processes, there exists some $\theta>0$, the time-averaged cost achieved by Algorithm \ref{alg:scheduling} is within a constant gap from $C^{\frac{(1+\theta)\Gamma}{\Gamma -w^{max}}}$, which is the offline minimum total cost when the total workload arrival rate to the federation is within $\frac{\Gamma -w^{max}}{(1+\theta)\Gamma}$ of the total workload service rate. i.e.,

\vspace{-4mm}
{\small\begin{align*}
& \lim_{R\to \infty} \frac{1}{R\Gamma} \sum_{n=0}^{R-1}\sum_{t=n\Gamma}^{(n+1)\Gamma - 1}\mathbb{E}[\sum_{j=1}^JC_j(t)] \\
& \le C^{\frac{(1+\theta)\Gamma}{\Gamma -w^{max}}} + \frac{B_1}{V} + \frac{B_2\Gamma}{V} + \frac{B_3}{\Gamma V} + \frac{B_4}{\Gamma}
\end{align*}
}
\noindent Here, $B_1, B_2, B_3, B_4$ are constants.
}

If $V \to \infty$, $\Gamma \to \infty$, $\frac{\Gamma}{V} \to \infty$ and $\theta$ scales down infinitely close to $0$, our algorithm achieves time-averaged cost infinitely close to the offline optimum. The proof of Theorem 3 is similar to our technique used in \cite{Profit_Single_Cloud} for proving the performance of extended Lyapunov optimization which takes jobs with varying workloads into consideration. The detail of the proof is in our technical report \cite{techreport2013}.

\section{A Double Auction Mechanism for One-Shot Cost Minimization} \label{sec:game}
In Sec.~\ref{sec:method}, our algorithm assumes a cooperative environment where individual cloud's workload and cost information are known for global cost minimization. We next design a double auction mechanism that elicits such truthful information, and guarantees global cost minimization in cases of selfish clouds. 




\vspace{-1mm}
\subsection{The Double Auction Model}
Each individual cloud in the federation is an agent in the double auction, whose strategies are buy-bids and sell-bids submitted to the auctioneer, a third party broker in the federated cloud.


\noindent \textbf{Individual Cloud's Utility. }
Let $P(i)$ and $P(l)$ denote cloud $i$'s proceeds from VM sales and charges paid for outsourcing a job $l$, respectively. Let $x_i^l$ be the allocation variable such that when $x_i^l =1$, the bid for job $l$ wins a bundle of $g^{k_l}$ VMs from cloud $i$ ( job $l$ can be migrated or newly scheduled to run in cloud $i$); when $x_i^l = 0$, the bid does not win a bundle of $g^{k_l}$ VMs from cloud $i$. A cloud can simultaneously win a set of atomic buy-bids. Its utility from winning a bid for a leftover job $l\in \widetilde{\mathcal{U}}_j$ is the reduction of its one-shot drift-plus-penalty through it, {\small$\mu(\mathbf{x^l}) = V\cdot\sum_{i=1}^J[(\beta_{h_l} - \lambda_{h_l}\cdot \phi^{k_l})g^{k_l} - P(l)].$
}

The utility from winning a bid for one new job $l\in \mathcal{U}_j^{k^*_j,max}$ 
is
{\small$\mu(\mathbf{x^l}) = V\cdot \sum_{i=1}^J[ \frac{(w^{k^*_j})^2q_j^{k^*_j}(t) + Z_j^{k^*_j}(t)}{V} - P(l) ].$
}

The utility of an individual cloud provider from selling VMs is {\small $\mu(\mathbf{x_i}) = V\cdot [P(i) - \sum_{j\in\mathcal{J}}\sum_{l\in \widetilde{\mathcal{U}}_j\cup\mathcal{U}_j^{k_j^*,max}}\beta_i g^{k_l}x_i^l].$
}

Now we can derive a cloud's valuation for a bundle of VMs:

\vspace{1mm}
\noindent {\em Valuation of VMs for leftover jobs: }
For leftover job $l \in \widetilde{\mathcal{U}}_j$, each job will request $g^{k_l}$ VMs. Its valuation for a bundle of $g^{k_l}$ VMs is $g^{k_l}\cdot[\beta_{h_l}(t)- \lambda_{h_l}\phi^{k_l}]$. Here $h_l$ is the cloud where leftover job $l$ is at the beginning of time slot $t$.

\vspace{1mm}
\noindent {\em Valuation of VMs for new jobs: }
For a new type-$k_j^*$ job from cloud $j$, its valuation for one bundle of $g^{k_j^*}$ VMs 
is $[(w^{k_j^*})^2q_j^{k_j^*}(t) + Z_j^{k_j^*}(t)]/V$.

\noindent \textbf{Bidding. } A cloud provider's buy-bid $(g_i^{k_l}, b^l)$ contains a request for $g^{k_l}$ co-located VMs in cloud $i$, for hosting a job $l$, and a bid price $b^l$.
The cloud provider can submit a number of {\em XOR'ed} VM bundle bids for each leftover job $l\in\widetilde{\mathcal{U}}_j$: $( g_1^{k_l}, b^l), \ldots ,( g_J^{k_l}, b^l)$, but at most one of them can win.
The cloud provider bids for multiple bundles for only one type of new jobs, type $k_j^* = argmax_k \frac{(w^k)^2q_j^k(t) + Z_j^k(t)}{g^k}$. The number of bundles the cloud provider wants to buy for its new jobs of type $k_j^*$ are as many as possible, because when the cost of one bundle for one new job is below the buy-bid, the cloud provider could increase its utility when obtaining it. We use {\small $U_j^{k_j^*,max} = |\mathcal{U}_j^{k_j^*,max}|$} to represent the number of bundles cloud provider $j$ bids for for its new type-$k_j^*$ jobs, $\mathcal{U}_j^{k_j^*,max}$ is the set of new jobs cloud provider $j$ submits bids for. The bid for one bundle is an {\em XOR'ed} bid of: $( g_1^{k_l}, b^l), \ldots ,( g_J^{k_l}, b^l)$. 

A cloud provider's sell-bid is simpler, consisting of an ask-price $s_j$ for one VM and the maximum number of supply $N_jH_j-\sum_{i\in\mathcal{J}}\sum_{l\in \widetilde{\mathcal{U}}_i} \sum_{s=1}^{g^k} I^l_{j,s}(t-1)$. 

\noindent \textbf{Winner Determination. } After collecting all the bids from cloud providers, the auctioneer solves the winner determination problem (WDP), to make decisions on allocation variables $x_i^l$'s to maximize the total surplus, {\em i.e.}, the total bid price for buying VMs minus the total ask price for selling VMs. 

\vspace{-6mm}
{\small\begin{align*}
\max & \quad \sum_{j\in\mathcal{J}}\sum_{l\in \widetilde{\mathcal{U}}_j\cup\mathcal{U}_j^{k_j^*,max}}\sum_{i=1}^J [b^l \cdot x_i^l - s_i\cdot g^{k_l}x_i^l] \nonumber \\
\end{align*}
}
\vspace*{-16mm}

{\small\begin{align*}
\mbox{s.t. } & \sum_{i=1}^J x_i^l \le 1, \forall l\in \widetilde{\mathcal{U}}_j\cup\mathcal{U}_j^{k_j^*,max}, 1\le j\le J; \nonumber \\
\end{align*}
}
\vspace*{-16mm}

{\footnotesize\begin{align}
& \sum_{j\in\mathcal{J}}\sum_{l\in \widetilde{\mathcal{U}}_j\cup\mathcal{U}_j^{k_j^*,max}} g^{k_l}x_i^l \le N_iH_i-\sum_{j\in \mathcal{J}}\sum_{l\in \widetilde{\mathcal{U}}_j} g^{k_l} I^l_{i,0}(t-1), \nonumber \\
& \quad\quad\quad\quad\quad\quad\quad 1\le i\le J; \nonumber
\end{align}
}

\vspace*{-12mm}

{\small\begin{align}
& x^l_i \in \{ 0,1 \}, \forall l\in \widetilde{\mathcal{U}}_j\cup\mathcal{U}_j^{k_j^*,max}.
\label{winner_determination}
\end{align}
}
\vspace{-3mm}

We point out the close connection between WDP (\ref{winner_determination}) and problem (\ref{migration_scheduling_allinone}) in the following claim. 

\noindent \textbf{Claim 4. } {\em The objective function of WDP (\ref{winner_determination}) equals {\small $-\frac{1}{V}\widetilde{\varphi_2}(t) + \sum_{j\in\mathcal{J}}\sum_{l\in \widetilde{\mathcal{U}}_j}g^{k_l}\beta_{h_l}(t)$ when $b^l = g^{k_l}\cdot[\beta_{h_l}(t)- \lambda_{h_l}\phi^{k_l}]$ for $l \in \widetilde{\mathcal{U}}_j$, $b^l = [(w^{k_j^*})^2q_j^{k_j^*}(t) + Z_j^{k_j^*}(t)]/V$ for $l\in \mathcal{U}_j^{k_j^*,max}$, $s_i = \beta_i(t)$, and $x_i^l$'s correspond to $I_{i,0}^l$'s as follows: for a new job $\mathcal{U}_j^{k_j^*}$, $x_i^l = I_{i,0}^l$; for a new job $l\in\mathcal{U}_j^{k_j^*,max} - \mathcal{U}_j^{k_j^*}$, $x_i^l = 0$; for a leftover job $l\in\widetilde{\mathcal{U}}_j$, $x_i^l = I_{i,0}^l, \forall i\neq h_l$; $x_{h_l}^l = 0$.} }

The details of the proof is in Appendix \ref{sec:appendix_c}.

Claim 4 shows that when each cloud provider bids truthfully for buying bundles of VMs for jobs and for selling VMs. We can derive the optimal allocation decisions $x_i^l$'s from the optimal solution to (\ref{migration_scheduling_allinone}), $I_{i,0}^l$'s, through the correspondence between them. Hence, Algorithm \ref{alg:scheduling} can compute an approximation result $APX$ no smaller than $LPR^* -\frac{C}{V}$, {\em i.e.}, $APX \ge LPR^* - \frac{C}{V}$, where $APX$ is the approximation solution under Algorithm \ref{alg:scheduling}, and $LPR^*$ is the optimal solution of the LP relaxation of WDP (\ref{winner_determination}). This can also be written as $APX \ge (1-\delta)\cdot LPR^*$, where $\delta = \frac{C}{V\cdot LPR^*}$.

Next, to elicit truth-telling from individual clouds, we design a randomized double auction that is truthfulness in expectation, employing Algorithm \ref{alg:scheduling} as a building block. 

\vspace{-1mm}
\subsection{A Truthful Randomized Double Auction Mechanism Design}

As shown in Algorithm \ref{alg:randomized_auction}, the randomized double auction includes four main steps. (i) We run the fractional VCG mechanism through solving the LP relaxation of the WDP, obtaining the fractional VCG allocation and charges/payments. (ii) We apply the LP duality based decomposition technique for decomposing the fractional VCG allocation, scaled by $(1-\delta)$, into a weighted combination of integral allocations. (iii) We randomly choose an integral allocation from the combination, with weights taken as probabilities. (iv) the fractional VCG payments are scaled down by a factor of $(1-\delta)$ to be the payments.

\begin{algorithm}[htb]
{\small
\caption{A Randomized Double Auction at time $t$}
\label{alg:randomized_auction}
\begin{algorithmic}[1]
\STATE Solve the LPR of (\ref{winner_determination}) to obtain solution $\mathbf{x^*}$, determine fractional payments $P^F(l)$ and $P^F(i)$ according to VCG mechanism.
\STATE Decompose the scaled down $\mathbf{x^*}$ into weighted integral solutions: $(1-\delta)\mathbf{x^*} = \sum_{z}\rho_z\mathbf{x(z)}$
 \STATE  \quad\quad\quad\quad - Solve primal-dual LPs using the ellipsoid method;
 \STATE \quad\quad\quad\quad  - employ Algorithm \ref{alg:scheduling} as the separation oracle
\STATE  Select $\mathbf{x(z)}$'s randomly with probability $\rho_z$;
\FOR {$l\in \widetilde{\mathcal{U}}_j\cup\mathcal{U}_j^{k_j^*,max}, 1\le j\le J$ }
    \STATE cloud provider $j$ pays $P^R(l) =  P^F(l) \cdot \frac{1}{\sum_{i=1}^Jx^{l*}_i}$ for winning a bid for its job $l$
\ENDFOR
\FOR {$1\le i\le J$}
    \STATE Cloud provider $i$ is paid {\small$P^R(i) = P^F(i) \cdot \sum_{j=1}^J\sum_{l\in \widetilde{\mathcal{U}}_j\cup\mathcal{U}_j^{k_j^*,max}}x^l_i(z)/\sum_{j=1}^J\sum_{l\in \widetilde{\mathcal{U}}_j\cup\mathcal{U}_j^{k_j^*,max}}x^{l*}_i$} for selling VMs under allocation $\mathbf{x(z)}$
\ENDFOR
\end{algorithmic}}
\end{algorithm}

\noindent \textbf{The Fractional VCG Pricing Mechanism. }
The LPR of the WDP can be solved efficiently. Let $\mathbf{x^*}$ denote its optimal (fractional) solution. The VCG payment for each winning bid for a job is $P^F(l) = $

\vspace{-4mm}
{\footnotesize\begin{align}
V(l) - \sum_{j\in\mathcal{J}}\sum_{i=1}^J[\sum_{l'\neq l, l'\in \widetilde{\mathcal{U}}_j\cup\mathcal{U}_j^{k_j^*,max}} b^{l'}\cdot x_i^{l'} -\sum_{l\in\widetilde{\mathcal{U}}_j\cup\mathcal{U}_j^{k_j^*,max}}s_i\cdot g^{k_l} x_i^l].
\end{align}
}
\vspace{-3mm}

\noindent where $V(l)$ is the solution of the LPR of (\ref{winner_determination}) when the cloud provider bids $0$ for a leftover job $l$ or one new job $l$.
The VCG payments to a cloud provider for provisioning VMs is $P^F(i) =$

\vspace{-4mm}
{\footnotesize\begin{align}
\sum_{j\in\mathcal{J}}\sum_{ l\in \widetilde{\mathcal{U}}_j\cup\mathcal{U}_j^{k_j^*,max}}[\sum_{i=1}^J b^l\cdot x_i^l  - \sum_{i'\neq i}s_{i'}\cdot g^{k_l}\cdot x_{i'}^l] - V(i).
\end{align}
}
\vspace{-4mm}

\noindent where $V(i)$ is the optimal solution of the LPR when available VMs at cloud $i$ is $0$.

\noindent \textbf{Decomposition of Fractional Solutions. }
The fractional VCG pricing mechanism achieves truthfulness, but its allocation is not practically applicable due to its fractional nature. Algorithm \ref{alg:scheduling} can compute an approximate integral solution to (\ref{winner_determination}). Let $APX$, $WDP^*$, $LPR^*$ denote the solution of WDP under Algorithm \ref{alg:scheduling}, the optimal solution of WDP, the optimal solution of the LPR, respectively. Algorithm \ref{alg:scheduling} achieves an approximate result that is at least $(1-\delta)$ times the LPR. The approximation ratio $\frac{1}{(1-\delta)}$ also verifies the integral gap of (\ref{winner_determination}). In fact, we have the following relation:

\vspace{-4mm}
{\small\begin{align*}
\mbox{Integrality gap} = \frac{LPR^*}{WDP^*} \le \frac{LPR^*}{APX} \le \frac{1}{(1-\delta)}
\end{align*}
}
\vspace{-4mm}

where the first inequality is due to $APX \le WDP^*$.

Applying the decomposition technique of Carr {\em et.al.}~\cite{Carr_RandomizedMetarounding} and Lavi {\em et.al.}~\cite{Lavi_RandomizedAuction}, we have the following result: Let $\mathcal{Z}$ be the set of all integral solutions of WDP. With the LPR of WDP (\ref{winner_determination}), and a $\frac{1}{(1-\delta)}$-approximation algorithm such as Algorithm \ref{alg:scheduling}, there is a polynomial algorithm that finds a polynomial number of integral solutions $\mathbf{x(1)}, \mathbf{x(2)}, \ldots, \mathbf{x(z)},\ldots, z\in\mathcal{Z}$ of WDP such that,

\vspace{-4mm}
{\small\begin{align*}
(1-\delta)\mathbf{x^*} = \sum_{z}\rho_z\mathbf{x(z)}
\end{align*}
}
\vspace{-4mm}

The polynomial number of integral solutions $\mathbf{x(z)}$ and their corresponding coefficients $\rho_z$ can be obtained by solving the following LP:

\vspace{-4mm}
{\small\begin{align*}
\min \quad &\sum_{z} \rho_z \\
\mbox{s.t. } & \sum_{z} \rho_z x^l_i(z) = (1-\delta)x_i^{l*}, l\in \widetilde{\mathcal{U}}_j\cup\mathcal{U}_j^{k_j^*,max}, 1\le i,j\le J; \\
& \sum_z \rho_z \ge 1; \\
& \rho_z \ge 0, \forall z \in \mathcal{Z}.
\end{align*}
}
\vspace{-4mm}

\noindent Its dual is:

\vspace{-4mm}
{\small\begin{align*}
\max \quad & (1-\delta)\sum_{j=1}^J\sum_{l\in \widetilde{\mathcal{U}}_j\cup\mathcal{U}_j^{k_j^*,max}}\sum_{i=1}^Jx_i^{l*} \cdot \nu_i^l + \theta \\
\mbox{s.t. } \quad & \sum_{j=1}^J\sum_{l\in \widetilde{\mathcal{U}}_j\cup\mathcal{U}_j^{k_j^*,max}}\sum_{i=1}^Jx_i^l(z) \cdot \nu_i^l + \theta \le 1, \forall z\in\mathcal{Z}; \\
& \nu_i^l \quad \mbox{unconstrained}, l\in \widetilde{\mathcal{U}}_j\cup\mathcal{U}_j^{k_j^*,max}, 1\le i,j\le J; \\
& \theta \ge 0.
\end{align*}
}
\vspace{-3mm}

The primal has an exponential number of variables, so we consider its dual. The dual has an exponential number of constraints and variables $\nu_i^l$'s and $\theta$. We can apply the ellipsoid method to solve the dual, with Algorithm \ref{alg:scheduling} used as a separation oracle.

Before illustrating the ellipsoid method, we show the following claim: 

\noindent \textbf{Claim 5. } {\em Let $\mathbf{\nu}=\{ \nu_i^l \}$ be any vector. $\nu_i^{l+} = \max \{0, \nu_i^l\}$. Given any integer solution $\mathbf{\widehat{x}}$ obtained from Algorithm \ref{alg:scheduling} for objective function {\small$\sum_{j=1}^J\sum_{l\in \widetilde{\mathcal{U}}_j\cup\mathcal{U}_j^{k_j^*,max}}\sum_{i=1}^Jx_i^{l} \cdot \nu_i^{l+}$}, one can obtain $\mathbf{x}(z)$ by letting $x_i^l(z) = \widehat{x}_i^l$ when $\nu_i^{l} \ge 0$ and $0$ otherwise. We have
{\small
$\sum_{j=1}^J\sum_{l\in \widetilde{\mathcal{U}}_j\cup\mathcal{U}_j^{k_j^*,max}}\sum_{i=1}^J x_i^{l}(z) \cdot \nu_i^l \\
\ge (1-\delta) \max_{\mathbf{x}\in \mathcal{P}}\sum_{j=1}^J\sum_{l\in \widetilde{\mathcal{U}}_j\cup\mathcal{U}_j^{k_j^*,max}}\sum_{i=1}^J x_i^{l}\cdot \nu_i^{l}$.
}
}

The details of the proof is in Appendix \ref{sec:appendix_d}.

The claim shows that for any vector of $\mathbf{\nu}$, we could find an integer solution $\mathbf{x}(z)$ which achieves the social welfare no smaller than $(1-\delta)$ times the optimal solution of the LPR. Based on this, we can prove that the optimal solution of the dual is $1$. Since $\nu_i^l = 0, \theta = 1$ is feasible, the optimal solution of the dual is at least $1$. Let $\nu_i^{l*}$, $\theta^*$ be given such that $(1-\delta)\sum_{j=1}^J\sum_{l\in \widetilde{\mathcal{U}}_j\cup\mathcal{U}_j^{k_j^*,max}}\sum_{i=1}^Jx_i^{l*} \cdot \nu_i^{l*} + \theta^* >1$. According to the claim, there exists a solution $\mathbf{x(z)}$ such that $\sum_{j=1}^J\sum_{l\in \widetilde{\mathcal{U}}_j\cup\mathcal{U}_j^{k_j^*,max}}\sum_{i=1}^Jx_i^l(z)\cdot \nu_i^{l*} > 1-\theta^*$. Hence, $\nu_i^{l*}$, $\theta^*$ violate the constraint $\sum_{j=1}^J\sum_{l\in \widetilde{\mathcal{U}}_j\cup\mathcal{U}_j^{k_j^*,max}}\sum_{i=1}^Jx_i^l(z) \cdot \nu_i^l + \theta \le 1$.

With the dual's optimal solution equal to $1$, we next illustrate how to apply Algorithm \ref{alg:scheduling} as the separation oracle that returns the separation hyperplane in the ellipsoid method. Let $\nu_i^{l*}$, $\theta^*$ denote the center of the current ellipsoid. When {\small$(1-\delta)\sum_{j=1}^J\sum_{l\in \widetilde{\mathcal{U}}_j\cup\mathcal{U}_j^{k_j^*,max}}\sum_{i=1}^Jx_i^{l*} \cdot \nu_i^{l*} + \theta^* < 1$}, we use the half space {\small$(1-\delta)\sum_{j=1}^J\sum_{l\in \widetilde{\mathcal{U}}_j\cup\mathcal{U}_j^{k_j^*,max}}\sum_{i=1}^Jx_i^{l*} \cdot \nu_i^{l*} + \theta^* \ge 1$} to cut the current ellipsoid; otherwise, we find the separation hyperplane using Algorithm \ref{alg:scheduling} as follows: Let $\nu_i^{l+} = \max \{0, \nu_i^{l*} \}$. Apply Algorithm \ref{alg:scheduling} to objective function {\small$\sum_{j=1}^J\sum_{l\in \widetilde{\mathcal{U}}_j\cup\mathcal{U}_j^{k_j^*,max}}\sum_{i=1}^Jx_i^{l} \cdot \nu_i^{l+}$} to get an integral solution $\widehat{\mathbf{x}}(z)$. Let $\mathbf{x}(z) = \widehat{\mathbf{x}}(z)$ if $\nu_i^{l*} \ge 0 $ and $0$ otherwise. According to the fact, we have {\small$\sum_{j=1}^J\sum_{l\in \widetilde{\mathcal{U}}_j\cup\mathcal{U}_j^{k_j^*,max}}\sum_{i=1}^Jx_i^{l}(z) \cdot \nu_i^{l*} \ge (1-\delta) \max_{\mathbf{x}\in \mathcal{P}}\sum_{j=1}^J\sum_{l\in \widetilde{\mathcal{U}}_j\cup\mathcal{U}_j^{k_j^*,max}}\sum_{i=1}^J x_i^{l}\cdot \nu_i^{l*} \ge 1- \theta^*$}, which can be used as the separation hyperplane.

\subsection{Properties of Randomized Double Auction Mechanism}

The fractional VCG double auction mechanism can give a truthful and individual rational mechanism with allocation $f^F(b^l, s_i) = \mathbf{x^*}$, the pricing scheme is $p^F(b^l, s_i)$: $(P^F(l)$,$P^F(i))$. The fractional solutions make it unpractical in real applications.

Our designed randomized double auction mechanism determines the allocation $f^R(b^l, s_i) = \mathbf{x(z)}$ with a probability $\rho_z$, $z\in\mathcal{Z}$. The pricing scheme is $p^R(b^l, s_i)$: $P^R(l) = 0$ if job $l$ does not win a bid, $P^R(l) =  P^F(l) \cdot \frac{1}{\sum_{i=1}^Jx^{l*}_i}$ if job $l$ wins a bid; $P^R(i) = 0$ if cloud $i$ does not sell VMs, $P^R(i) = P^F(i) \cdot \frac{\sum_{j=1}^J\sum_{l\in \widetilde{\mathcal{U}}_j\cup\mathcal{U}_j^{k_j^*,max}}x^l_i(z)}{\sum_{j=1}^J\sum_{l\in \widetilde{\mathcal{U}}_j\cup\mathcal{U}_j^{k_j^*,max}}x^{l*}_i}$ if cloud $i$ sells VMs and the randomly determined allocation is $\mathbf{x(z)}$.
The price is set this way to let the expectation of job $l$'s price equals to $(1-\delta)\cdot P^F(l)$ and the expectation of cloud $i$'s revenue equals to $(1-\delta)\cdot P^F(i)$.

\noindent \textbf{Definition (Truthfulness in expectation)} {\em A randomized double auction mechanism $(f^R,p^R)$ is truthful in expectation if for any buyer $l$ or seller $i$, $\mathbb{E}[\mu(f^R(\tilde{b}^l, b^{-l}, s_{i}))] \ge \mathbb{E}[\mu(f^R(b^{l'}, b^{-l}, s_{i}))]$, and $\mathbb{E}[\mu(f^R(b^l, \tilde{s}_{i}, s_{-i}))] \ge \mathbb{E}[\mu(f^R(b^l, s'_{i}, s_{i}))]$, $b^{-l}$ represents any bids from other buyers except buyer $l$, $s_{-i}$ represents any bids from other sellers except seller $i$, $\tilde{b}^l$ and $\tilde{s}_{i}$ are the true valuations of buyer $l$ and seller $i$, $b^{l'}$ and $s'_{i}$ are any other bids of buyer $l$ and seller $i$.
}

We have the following theorem for the properties of our designed randomized double auction mechanism.

\noindent \textbf{Theorem 6. } {\em The randomized double auction mechanism in Algorithm \ref{alg:randomized_auction} is truthful in expectation, individual rational, and $\frac{1}{1-\delta}$-approximate to the WDP in (\ref{winner_determination}). 
}

\noindent {\em Proof sketch:} We can define a deterministic support mechanism for the randomized double auction mechanism. The allocation of the deterministic support mechanism is $f^D(b^l,s_i) = (1-\delta)\cdot \mathbf{x}^* = \sum_{z} \rho_z \mathbf{x}(z)$, the pricing scheme is $p^D(b^l,s_i)$: $P^D(l) = (1-\delta) P^F(l) = \mathbb{E}[P^R(l)]$, $P^D(i) = (1-\delta) P^F(i) = \mathbb{E}[P^R(i)]$. It is easy to prove the truthfulness, individual rationality and $\frac{1}{1-\delta}$-approximation of the deterministic support mechanism based on the VCG mechanism. We can prove the randomized double auction mechanism preserves truthfulness, individual rationality, and $\frac{1}{1-\delta}$-approximation of the support mechanism.

Theorem 6 shows that the randomized double auction mechanism can achieve the same approximation ratio with Algorithm \ref{alg:scheduling}. Hence, according to Theorem 3, in the long term, it achieves the time-averaged cost that is within a constant gap from the offline optimum.

\section{Simulations} \label{sec:simulations}


We evaluate our online resource trading algorithm among $3$ IaaS clouds located in $3$ different regions in North America (N.Virginia, Oregon, Northern California). By default, each cloud provider's data center has $10^3$ servers. Each server can host 10 VMs. We extract the spot prices of medium instances in different regions of Amazon EC2 \cite{AMAZONEC2} as the costs of providing one VM.
We use the price for data transfer out from Amazon EC2 to Internet, i.e., $\$0.12$ per GB, to calculate the VM migration cost. The data needed to transfer for migrating one VM is randomly selected in range $0$-$5$ GB.
One time slot is $1$ hour. The time length of a job and the number of VMs in a job are both between $[1,5]$.  We extract hourly job arrival patterns from the Google cluster trace \cite{clusterdata:Reiss2011} as our input. The number of job arrivals in each hour to an individual cloud is set according to the cumulated job requests of each type submitted to the Google cluster during that hour.

\vspace{-1mm}
\subsection{Clouds' Cost When They Trade or Not}

We run Algorithm \ref{alg:scheduling} for each cloud separately to simulate one cloud's operation without VM resource trading, and we run Algorithm \ref{alg:randomized_auction} to simulate $3$ clouds' operation with VM trading, both under different values of $V$ and $\Gamma$. We compare the time-averaged cost of each individual cloud over $240$ hours under no trading and trading scenarios.

\vspace{-4mm}
\begin{figure}[!htbp]
\centering
\includegraphics[width=0.35\textwidth]{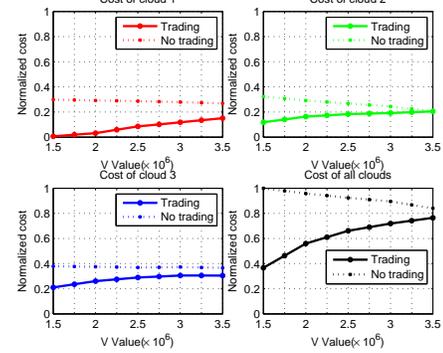}
\vspace{-3mm}
\caption{Clouds' cost when they trade or do not trade for different $V$ values.}
\label{fig:dynamicpricing}
\vspace{-4mm}
\end{figure}

\begin{figure}[!htbp]
\centering
\includegraphics[width=0.35\textwidth]{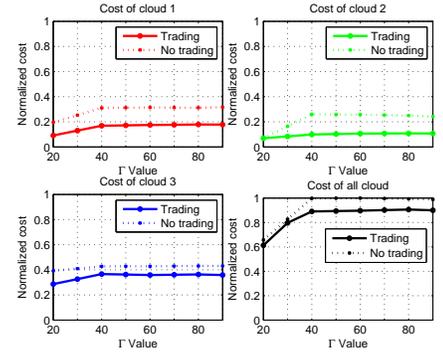}
\vspace{-3mm}
\caption{Clouds' cost when they trade or do not trade for different $\Gamma$ values.}
\label{fig:comparison}
\vspace{-4mm}
\end{figure}

The cost for each individual cloud and the total cost are depicted in Fig.~\ref{fig:dynamicpricing} and Fig.~\ref{fig:comparison}. Fig.~\ref{fig:dynamicpricing} shows how the cost changes with parameter $V$, here $\Gamma$ is set as $16w^{max}$. Parameter $V$ represents the tradeoff ratio between drift {\em representing job queue backlog} and penalty {\em i.e. cost} in the one slot drift-plus-penalty minimization.When $V$ increases, the cost when they do not trade decreases, but the cost when they trade increases. This is because the valuation for VMs is inversely proportional to $V$. Under the same trading volume, a larger $V$ means fewer income from trading. Fig.~\ref{fig:comparison} shows how the cost changes with parameter $\Gamma$. As $\Gamma$ increases, {\em i.e. the length of one time frame in which scheduled jobs should be completed}, the cost will increase to a stable value, because the limiting power of the time frame on job scheduling becomes weaker as the length of time frame is larger.

\vspace{-1mm}
\subsection{Ratio of Active Servers}
\begin{figure}[!htbp]
\centering
\includegraphics[width=0.26\textwidth]{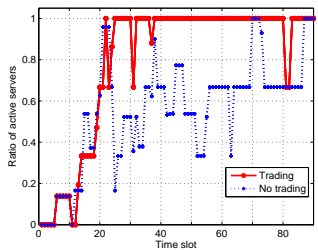}
\caption{Ratio of active servers.}
\label{fig:activeratio}
\vspace{-4mm}
\end{figure}

Fig.~\ref{fig:activeratio} shows the ratio of total active servers in all clouds when they trade or do not trade. The ratio of servers' utilization under trading is higher than that under no trading. This could be to the contrary of reducing cost at the first sight. However, while clouds trade their idle resources to serve others' high-value jobs and earn income, they can achieve better cost reduction than letting resources idle.

\vspace{-1mm}

\vspace{0mm}
\subsection{Cost vs.~Job Load}
\begin{figure}[!htbp]
\centering
\includegraphics[width=0.26\textwidth]{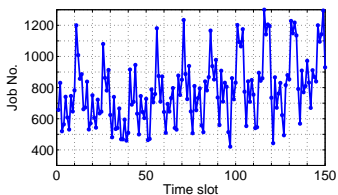}
\vspace{-3mm}
\caption{Cloud $3$'s job pattern.}
\label{fig:job_pattern}
\vspace{-4mm}
\end{figure}

\begin{figure}[!htbp]
\centering
\includegraphics[width=0.26\textwidth]{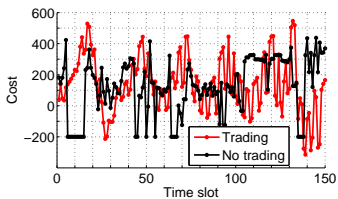}
\vspace{-3mm}
\caption{Cloud $3$'s cost pattern.}
\label{fig:cost_pattern}
\vspace{-4mm}
\end{figure}

Fig.~\ref{fig:job_pattern} shows job arrivals in cloud $3$. Fig.~\ref{fig:cost_pattern} shows the corresponding cost under trading and no trading. The pattern of cost under trading matches the job pattern more closely. The dynamic range of the costs when they trade is larger than that when they do not trade, as when the job load is small, idle servers can be sold to other clouds to obtain income to compensate the cost.

\vspace{-4mm}

\vspace{-1mm}
\subsection{Average Delay of Jobs}
\begin{figure}[!htbp]
\centering
\includegraphics[width=0.26\textwidth]{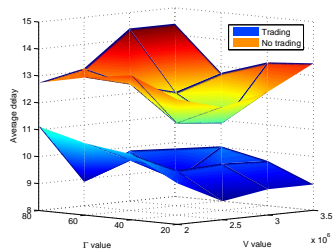}
\caption{Comparison of average delay of jobs.}
\label{fig:average_delay}
\vspace{-4mm}
\end{figure}

We measure the average service response delay of jobs under different $V$ and $\Gamma$. Fig.~\ref{fig:average_delay} compares the average delay of jobs when the clouds trade or not. The average delay actually experienced by jobs is smaller with trading than without. The reason is that dynamic costs in different data centers increase the chances of jobs being scheduled at a time slot. As long as one data center has low prices, jobs can be scheduled to execute there.

\section{Conclusions}
\label{sec:conclusion}
We formulated a global cost minimization problem with inter-cloud job
scheduling and migration for exploiting the temporal and spatial
diversities in operational cost among federated clouds. Lyapunov optimization theory
is applied for translating such long-term optimization into a one-shot minimization
problem. An effective approximation algorithm is designed, and is translated into a truthful randomized double
auction through a recent LP-based decomposition technique, with the same 
guarantee in approximation ratio. 

\bibliographystyle{IEEEtran}
\bibliography{Document_Management}

\appendices
\label{sec:appendix}

\section{Derivation of one-shot drift-plus penalty} \label{sec:appendix_a}

The one-shot drift for cloud provider $j$ is

{\small\begin{align*}
& L(\mathbf{\Theta}_j(t+1)) - L(\mathbf{\Theta}_j(t)) \\
&=  \frac{1}{2}\sum_{k=1}^K[(w^kq_j^k(t+1))^2 + Z_j^k(t+1)^2] \\
& - \frac{1}{2}\sum_{k=1}^K[(w^kq_j^k(t))^2 + Z_j^k(t)^2] \\
& = \frac{1}{2}\sum_{k=1}^K[(w^k)^2(q_j^k(t+1)^2 - q_j^k(t)^2) + Z_j^k(t+1)^2 - Z_j^k(t)^2 ] \\
& \le B + \sum_{k=1}^K[(w^k)^2q_j^k(t)(A_j^k(t) - U_j^k(t) - G_j^k(t)) \\
& + Z_j^k(\epsilon_j^k - U_j^k(t) - G_j^k(t))].
\end{align*}
}
Here $B = \frac{1}{2}\sum_{k=1}^K\{(w^k)^2(A^{max})^2 +(\epsilon_j^k)^2 + [(w^k)^2+1][U_j^{k,max} + G_j^{k,max}]^2 \}$.

The inequality is attributed to

{\small\begin{align*}
& q_j^k(t+1)^2 - q_j^k(t)^2 \\
&\le  [q_j^k(t) - U_j^k(t) - G_j^k(t)]^2 +A_j^k(t)^2 \\
& + 2A_j^k(t)\cdot \max\{ q_j^k(t) - U_j^k(t) - G_j^k(t)\} -q_j^k(t)^2\\
& \le  [U_j^k(t)+G_j^k(t)]^2 +A_j^k(t)^2 \\
& -2q_j^k(t)[U_j^k(t) + G_j^k(t)] + 2A_j^k(t)q_j^k(t) \\
& \le  [ U_j^{k,max}+G_j^{k,max} ]^2 + (A_j^{max})^2 \\
& + 2q_j^k(t)[A_j^k(t) - U_j^k(t) - G_j^k(t)],
\end{align*}
}
and

{\small\begin{align*}
& Z_j^k(t+1)^2 - Z_j^k(t)^2 \\
& \le [Z_j^k(t) + \epsilon_j^k - U_j^k(t)-G_j^k(t)]^2 -Z_j^k(t)^2 \\
& \le (\epsilon_j^k)^2 + [U_j^{k,max}+G_j^{k,max}]^2 \\
& + 2Z_j^k(t)[\epsilon_j^k - U_j^k(t)-G_j^k(t)].
\end{align*}
}


The one-shot drift for all clouds is

{\small\begin{align*}
& \Delta(\mathbf{\Theta}(t)) = \sum_{j=1}^J \mathbb{E}\{L(\mathbf{\Theta}_j(t+1)) - L(\mathbf{\Theta}_j(t))|\mathbf{\Theta}(t)\} \\
\le & B\cdot J + \sum_{j=1}^J\sum_{k=1}^K(w^k)^2q_j^k(t)\left\{\mathbb{E}[A_j^k(t)] - U_j^k(t) - G_j^k(t)\right\} \\
 & + \sum_{j=1}^J\sum_{k=1}^K Z_j^k\cdot[\epsilon_j^k - U_j^k(t) - G_j^k(t)],
\end{align*}
}
where $B$ is a constant.


The one-shot {\em drift-plus-penalty} is

{\small\begin{align*}
& \Delta(\mathbf{\Theta}(t)) + V\cdot \sum_{j=1}^J [\sum_{l\in \cup\widetilde{\mathcal{U}}_j}c^l + \sum_{l\in \cup_{\mathcal{K}}\mathcal{U}_j^k}c^l+ \sum_{k=1}^K\alpha^k_j G^k_j(t)] \\
\le & B\cdot J + \sum_{j=1}^J\sum_{k=1}^K(w^k)^2q_j^k(t)\left\{\mathbb{E}[A_j^k(t)] - U_j^k(t) - G_j^k(t)\right\} \\
 & + \sum_{j=1}^J\sum_{k=1}^K Z_j^k\cdot[\epsilon_j^k - U_j^k(t) - G_j^k(t)] \\
 & + V\cdot \sum_{j=1}^J [\sum_{l\in \cup\widetilde{\mathcal{U}}_j}c^l + \sum_{l\in \cup_{\mathcal{K}}\mathcal{U}_j^k}c^l+ \sum_{k=1}^K\alpha^k_j G^k_j(t)] \\
= & B\cdot J + \sum_{j=1}^J\sum_{k=1}^K(w^k)^2q_j^k(t)\mathbb{E}[A_j^k(t)] + \sum_{j=1}^J\sum_{k=1}^K Z_j^k\cdot\epsilon_j^k \\
& + \sum_{j=1}^J\sum_{k=1}^K G_j^k(t)[V\alpha^k_j -(w^k)^2q_j^k(t)-Z_j^k] \\
& + V\cdot \sum_{j=1}^J\sum_{l\in \cup_{\mathcal{K}}\mathcal{U}_j^k}c^l - \sum_{j=1}^J\sum_{k=1}^K U_j^k(t)[(w^k)^2q_j^k(t)+Z_j^k] \\
& + V\cdot \sum_{j=1}^J \sum_{l\in \cup\widetilde{\mathcal{U}}_j}c^l.
\end{align*}
}
Hence, the minimization of one-shot {\em drift-plus-penalty} is to minimize $\varphi_1(t) + \varphi_2(t)$,
where

\vspace{-4mm}
{\small\begin{align*}
\varphi_1(t) = & \sum_{j=1}^J\sum_{k=1}^K[ V\alpha_j^k - (w^k)^2q_j^k(t)-Z_j^k(t)]G_j^k(t), \\
\varphi_2(t) = & \sum_{l\in \cup_{\mathcal{J},\mathcal{K}}\mathcal{U}_j^k}Vc^l - \sum_{j=1}^J\sum_{k=1}^K [(w^k)^2q_j^k(t) + Z_j^k(t)]U_j^k(t) \\
& + \sum_{l\in \cup_{\mathcal{J}}\widetilde{\mathcal{U}}_j}Vc^l.
\end{align*}
\vspace{-3mm}
}

\section{Proof of Theorem 1}
\label{sec:appendix_b}

Let us first prove $\widetilde{\varphi_2}^*(t) \le \varphi_2^*(t)$, where $\widetilde{\varphi_2}^*(t)$ is the optimal solution of the LP Relaxation (LPR) of (\ref{migration_scheduling_allinone}), and $\varphi_2^*(t)$ is the optimal solution of (\ref{migration_scheduling}).

The expression for $\varphi_2(t)$ can be rearranged as follows,

{\small\begin{align*}
\varphi_2(t) = & \sum_{l\in \cup_{\mathcal{J},\mathcal{K}}\mathcal{U}_j^k}Vc^l - \sum_{j=1}^J\sum_{k=1}^K [(w^k)^2q_j^k(t) + Z_j^k(t)]U_j^k(t) \\
& + \sum_{l\in \cup_{\mathcal{J}}\widetilde{\mathcal{U}}_j}Vc^l \\
 = & \sum_{l\in \cup_{\mathcal{J},\mathcal{K}}\mathcal{U}_j^k}V\{ \sum_{i=1}^J\beta_i(t)\sum_{s=1}^{g^k}I^l_{i,s}(t) \\
 & + \sum_{h\neq i}\sum_{i=1}^J\lambda_h[I^l(t)\cdot T^l\cdot I^l(t)']_{h,i}\} \\
 & + \sum_{l\in \cup_{\mathcal{J}}\widetilde{\mathcal{U}}_j}V\{ \sum_{i=1}^J\beta_i(t)\sum_{s=1}^{g^k}I^l_{i,s}(t) \\
 & + \sum_{h\neq i}\sum_{i=1}^J \lambda_{h} [I^l(t)\cdot T^l\cdot I^l(t)']_{h,i}  \\
& + \sum_{h\neq i}\sum_{i=1}^J \lambda_{h}\sum_{s=1}^{g^k}I^l_{h,s}(t-1) \cdot I^l_{i,s}(t)\cdot \phi^k \} \\
& - \sum_{j=1}^J\sum_{k=1}^K [(w^k)^2q_j^k(t) + Z_j^k(t)]U_j^k(t) \\
= & \sum_{l\in \cup_{\mathcal{J},\mathcal{K}}\mathcal{U}_j^k}V\{ \sum_{i=1}^J\beta_i(t)\sum_{s=1}^{g^k}I^l_{i,s}(t) \} \\
& - \sum_{j=1}^J\sum_{k=1}^K [(w^k)^2q_j^k(t) + Z_j^k(t)]U_j^k(t) \\
& + \sum_{l\in \cup_{\mathcal{J}}\widetilde{\mathcal{U}}_j}V\{ \sum_{i=1}^J\beta_i(t)\sum_{s=1}^{g^k}I^l_{i,s}(t) \\
& + \sum_{h\neq i}\sum_{i=1}^J \lambda_{h}\sum_{s=1}^{g^k}I^l_{h,s}(t-1) \cdot I^l_{i,s}(t)\cdot \phi^k \} \\
& + \sum_{l\in \cup_{\mathcal{J},\mathcal{K}}\mathcal{U}_j^k}V\sum_{h\neq i}\sum_{i=1}^J\lambda_h[I^l(t)\cdot T^l\cdot I^l(t)']_{h,i} \\
& + \sum_{l\in \cup_{\mathcal{J}}\widetilde{\mathcal{U}}_j}V\sum_{h\neq i}\sum_{i=1}^J \lambda_{h} [I^l(t)\cdot T^l\cdot I^l(t)']_{h,i} \\
\ge & \sum_{l\in \cup_{\mathcal{J},\mathcal{K}}\mathcal{U}_j^k}V\{ \sum_{i=1}^J\beta_i(t)\sum_{s=1}^{g^k}I^l_{i,s}(t) \} \\
& - \sum_{j=1}^J\sum_{k=1}^K [(w^k)^2q_j^k(t) + Z_j^k(t)]U_j^k(t) \\
& + \sum_{l\in \cup_{\mathcal{J}}\widetilde{\mathcal{U}}_j}V\{ \sum_{i=1}^J\beta_i(t)\sum_{s=1}^{g^k}I^l_{i,s}(t) \\
& + \sum_{h\neq i}\sum_{i=1}^J \lambda_{h}\sum_{s=1}^{g^k}I^l_{h,s}(t-1) \cdot I^l_{i,s}(t)\cdot \phi^k \} \\
= & \sum_{l\in \cup_{\mathcal{J},\mathcal{K}}\mathcal{U}_j^k}\sum_{i=1}^J \{ V\beta_i(t)-  \frac{(w^k)^2q_j^k(t) + Z_j^k(t)}{g^k} \}\sum_{s=1}^{g^k}I^l_{i,s}(t)  \\
& + \sum_{l\in \cup_{\mathcal{J}}\widetilde{\mathcal{U}}_j}\sum_{i=1}^J \{ V\beta_i(t) + V\sum_{h\neq i} \lambda_{h}I^l_{h,s}(t-1) \cdot \phi^k\}\sum_{s=1}^{g^k}I^l_{i,s}(t).
\end{align*}
}

The inequality is because $\varphi_2(t)$ includes the cost for the inter-data center traffic among VMs of the same job located in different data centers, while the RHS of the inequality,
{\small\begin{align}
& \sum_{l\in \cup_{\mathcal{J},\mathcal{K}}\mathcal{U}_j^k}\sum_{i=1}^J \{ V\beta_i(t)-  \frac{(w^k)^2q_j^k(t) + Z_j^k(t)}{g^k} \}\sum_{s=1}^{g^k}I^l_{i,s}(t)  \nonumber\\
& + \sum_{l\in \cup_{\mathcal{J}}\widetilde{\mathcal{U}}_j}\sum_{i=1}^J \{ V\beta_i(t) + V\sum_{h\neq i} \lambda_{h}I^l_{h,s}(t-1) \cdot \phi^k\}\sum_{s=1}^{g^k}I^l_{i,s}(t), \label{intermediate}
\end{align}
}

\noindent does not take the cost for the inter-data center traffic among VMs of the same job into consideration.

When $ I_{i,0}^l(t) = \frac{\sum_{s=1}^{g^k}I_{i,s}^l(t)}{g^k} $,  $\widetilde{\varphi_2}(t)$ is equal to (\ref{intermediate}). Hence, for any $I^l_{i,s}(t)$, we have the optimal solution of the LP Relaxation of (\ref{migration_scheduling_allinone}), $\widetilde{\varphi_2}^*(t)$

{\small\begin{align*}
\le & \sum_{l\in \cup_{\mathcal{J},\mathcal{K}}\mathcal{U}_j^k}\sum_{i=1}^J \{ V\beta_i(t)-  \frac{(w^k)^2q_j^k(t) + Z_j^k(t)}{g^k} \}\sum_{s=1}^{g^k}I^l_{i,s}(t)  \nonumber\\
& + \sum_{l\in \cup_{\mathcal{J}}\widetilde{\mathcal{U}}_j}\sum_{i=1}^J \{ V\beta_i(t) + V\sum_{h\neq i} \lambda_{h}I^l_{h,s}(t-1) \cdot \phi^k\}\sum_{s=1}^{g^k}I^l_{i,s}(t).
\end{align*}
}

So, $\widetilde{\varphi_2}^*(t) \le \varphi_2^*(t)$.

We next prove that $\widetilde{\varphi_2}^{(a)}(t) \le \widetilde{\varphi_2}^*(t) + C$, where $C = VJg^{max}\cdot \max\{ \frac{\alpha^{max}}{g^{min}} - \beta_{min} ,\beta_{max}+ \lambda_{max}\phi_{max} - \beta_{min} -\lambda_{min}\phi_{min}\}$.

Let $\overline{I_{i,0}^l}(t)$ denote VM placement of jobs under the optimal solution of the LP Relaxation (LPR) of (\ref{migration_scheduling_allinone}).

{\small\begin{align*}
\widetilde{\varphi_2}^*(t) = & \sum_{l\in \cup_{\mathcal{J},\mathcal{K}}\mathcal{U}^k_j}\sum_{i=1}^J[ V\beta_i(t) - \frac{(w^k)^2q_j^k(t) + Z_j^k(t)}{g^k} ]g^k \overline{I^l_{i,0}}(t) \\
& + \sum_{l\in \cup_{\mathcal{J}}\widetilde{\mathcal{U}}_j} \sum_{i=1}^J [V\beta_i(t) + V\sum_{h\neq i} \lambda_{h}I^l_{h,0}(t-1) \cdot \phi^k ]g^k \overline{I^l_{i,0}}(t) \\
\ge_{(a)} & \sum_{i\in\mathcal{D}_L}[ V\beta_i(t) - \frac{(w^{k^*})^2q_{j^*}^{k^*}(t) + Z_{j^*}^{k^*}(t)}{g^{k^*}} ] \\
& \cdot [N_iH_i - \sum_{l\in\cup_{\mathcal{J}}\widetilde{\mathcal{U}}_j} g^k \overline{I^l_{i,0}}(t)] \\
& + \sum_{l\in \cup_{\mathcal{J}}\widetilde{\mathcal{U}}_j} \sum_{i=1}^J [V\beta_i(t) + V\sum_{h\neq i} \lambda_{h}I^l_{h,0}(t-1) \cdot \phi^k ]g^k \overline{I^l_{i,0}}(t) \\
= & \sum_{i\in\mathcal{D}_L}[ V\beta_i(t) - \frac{(w^{k^*})^2q_{j^*}^{k^*}(t) + Z_{j^*}^{k^*}(t)}{g^{k^*}} ]\cdot N_iH_i \\
& + \sum_{l\in \cup_{\mathcal{J}}\widetilde{\mathcal{U}}_j}\sum_{i\in\mathcal{D}_L}[\frac{(w^{k^*})^2q_{j^*}^{k^*}(t) + Z_{j^*}^{k^*}(t)}{g^{k^*}} \\
& + V\sum_{h\neq i} \lambda_{h}I^l_{h,0}(t-1) \cdot \phi^k ]  \cdot g^k \overline{I^l_{i,0}}(t) \\
& + \sum_{l\in \cup_{\mathcal{J}}\widetilde{\mathcal{U}}_j} \sum_{i\in\mathcal{D}_H} [V\beta_i(t) + V\sum_{h\neq i} \lambda_{h}I^l_{h,0}(t-1) \cdot \phi^k ]g^k \overline{I^l_{i,0}}(t).
\end{align*}
}

The inequality $(a)$ is because $\widetilde{\varphi_2}(t)$ has a lower bound when all its idle VMs in data centers $\mathcal{D}_L$ after leftover job migration are used to run new jobs of type-$k^*$ in cloud $j^*$, {\em i.e.}, the type of jobs with the largest value $\frac{(w^k)^2q_j^k(t) + Z_j^k(t)}{g^k}$.

Let $I_{i,0}^{l}(t)^{(a)}$ denote VM placement of jobs under Algorithm \ref{alg:scheduling}. Algorithm \ref{alg:scheduling} only schedules new jobs of type-$k^*$ in cloud $j^*$ to idle VMs in data centers $\mathcal{D}_L$ after leftover job migration. We have $\sum_{l\in \mathcal{U}^{k^*}_{j^*}}g^{k^*} I^l_{i,0}(t)^{(a)} = N_iH_i - \sum_{l\in \cup_{\mathcal{J}}\widetilde{\mathcal{U}}_j} g^k I^l_{i,0}(t)^{(a)} - g^{r_i}$, where $g^{r_i}$ is the number of idle VMs in data center $i$ that are not enough to host a type-$k^*$ job in cloud $j^*$, and $g^{r_i} < g^{k^*}$.

{\small\begin{align*}
& \widetilde{\varphi_2}^{(a)}(t) \\
= &  \sum_{l\in \mathcal{U}^{k^*}_{j^*}}\sum_{i\in \mathcal{D}_L}[ V\beta_i(t) - \frac{(w^{k^*})^2q_{j^*}^{k^*}(t) + Z_{j^*}^{k^*}(t)}{g^{k^*}} ]g^{k^*} I^l_{i,0}(t)^{(a)} \\
& + \sum_{l\in \cup_{\mathcal{J}}\widetilde{\mathcal{U}}_j} \sum_{i=1}^J [V\beta_i(t) + V\sum_{h\neq i} \lambda_{h}I^l_{h,0}(t-1) \cdot \phi^k ]g^k I^l_{i,0}(t)^{(a)} \\
= & \sum_{i\in \mathcal{D}_L}[ V\beta_i(t) - \frac{(w^{k^*})^2q_{j^*}^{k^*}(t) + Z_{j^*}^{k^*}(t)}{g^{k^*}} ] \\
& \cdot[N_iH_i - \sum_{l\in \cup_{\mathcal{J}}\widetilde{\mathcal{U}}_j} g^k I^l_{i,0}(t)^{(a)} - g^{r_i}] \\
& + \sum_{l\in \cup_{\mathcal{J}}\widetilde{\mathcal{U}}_j} \sum_{i=1}^J [V\beta_i(t) + V\sum_{h\neq i} \lambda_{h}I^l_{h,0}(t-1) \cdot \phi^k ]g^k I^l_{i,0}(t)^{(a)} \\
= & \sum_{i\in \mathcal{D}_L}[ V\beta_i(t) - \frac{(w^{k^*})^2q_{j^*}^{k^*}(t) + Z_{j^*}^{k^*}(t)}{g^{k^*}} ]N_iH_i \\
& - \sum_{i\in \mathcal{D}_L}[ V\beta_i(t) - \frac{(w^{k^*})^2q_{j^*}^{k^*}(t) + Z_{j^*}^{k^*}(t)}{g^{k^*}} ]g^{r_i} \\
& + \sum_{l\in \cup_{\mathcal{J}}\widetilde{\mathcal{U}}_j}\sum_{i\in\mathcal{D}_L}[\frac{(w^{k^*})^2q_{j^*}^{k^*}(t)+ Z_{j^*}^{k^*}(t)}{g^{k^*}} \\
& + V\sum_{h\neq i} \lambda_{h}I^l_{h,0}(t-1) \cdot \phi^k ]g^k I^l_{i,0}(t)^{(a)} \\
& + \sum_{l\in \cup_{\mathcal{J}}\widetilde{\mathcal{U}}_j} \sum_{i\in\mathcal{D}_H} [V\beta_i(t) + V\sum_{h\neq i} \lambda_{h}I^l_{h,0}(t-1) \cdot \phi^k ]g^k I^l_{i,0}(t)^{(a)}.
\end{align*}
}
The above expression currently only includes variables of VM placement of leftover jobs. According to Algorithm \ref{alg:scheduling}, the leftover jobs in the data center with the highest cost are migrated first to the available data center with the lowest cost. After the leftover job migration, there are two cases: (1) All leftover jobs are in or migrated to data centers in set $\mathcal{D}_L$. (2) There are leftover jobs in data center set $\mathcal{D}_H$.

Let us analyze the value of $\widetilde{\varphi_2}^{(a)}(t)$ under these two cases respectively.

(1)  All leftover jobs are in or migrated to data centers in $\mathcal{D}_L$:
In this case, $\sum_{l\in \cup_{\mathcal{J}}\widetilde{\mathcal{U}}_j} \sum_{i\in\mathcal{D}_H} [V\beta_i(t) + V\sum_{h\neq i} \lambda_{h}I^l_{h,0}(t-1) \cdot \phi^k ]g^k I^l_{i,0}(t)^{(a)} = \sum_{l\in \cup_{\mathcal{J}}\widetilde{\mathcal{U}}_j} \sum_{i\in\mathcal{D}_H} [V\beta_i(t) + V\sum_{h\neq i} \lambda_{h}I^l_{h,0}(t-1) \cdot \phi^k ]g^k \overline{I^l_{i,0}}(t) = 0$, and $\sum_{l\in \cup_{\mathcal{J}}\widetilde{\mathcal{U}}_j}\sum_{i\in\mathcal{D}_L}[\frac{(w^{k^*})^2q_{j^*}^{k^*}(t)+ Z_{j^*}^{k^*}(t)}{g^{k^*}}  + V\sum_{h\neq i} \lambda_{h}I^l_{h,0}(t-1) \cdot \phi^k ]g^k I^l_{i,0}(t)^{(a)} = \sum_{l\in \cup_{\mathcal{J}}\widetilde{\mathcal{U}}_j}\sum_{i\in\mathcal{D}_L}[\frac{(w^{k^*})^2q_{j^*}^{k^*}(t) + Z_{j^*}^{k^*}(t)}{g^{k^*}} + V\sum_{h\neq i} \lambda_{h}I^l_{h,0}(t-1) \cdot \phi^k ]  \cdot g^k \overline{I^l_{i,0}}(t) $. We have the following relationship between $\widetilde{\varphi_2}^{(a)}(t)$ and $\widetilde{\varphi_2}^*(t)$,

{\small\begin{align*}
\widetilde{\varphi_2}^{(a)}(t) = &  \widetilde{\varphi_2}^*(t) - \sum_{i\in \mathcal{D}_L}[ V\beta_i(t) - \frac{(w^{k^*})^2q_{j^*}^{k^*}(t) + Z_{j^*}^{k^*}(t)}{g^{k^*}} ]g^{r_i} \\
\le & \widetilde{\varphi_2}^*(t) + VJg^{max}[\frac{(w^{k^*})^2q_{j^*}^{k^*}(t) + Z_{j^*}^{k^*}(t)}{Vg^{k^*}} - \beta_i(t)] \\
\le_{(b)} & \widetilde{\varphi_2}^*(t) + VJg^{max}[ \frac{\alpha^{max}}{g^{min}} - \beta_{min}].
\end{align*}
}

The inequality $(b)$ is because $\alpha_j^k$'s are large enough to guarantee there is no job dropping, and the value $\frac{(w^{k^*})^2q_{j^*}^{k^*}(t) + Z_{j^*}^{k^*}(t)}{Vg^{k^*}}$ is bounded by $\frac{\alpha_{j^*}^{k^*}}{g^{k^*}} $.

(2) There are leftover jobs in data centers belonging to $\mathcal{D}_H$: In this case, $g^{r_i}$ is the number of remaining idle VMs in data center $i\in\mathcal{D}_L$ that are not enough for hosting an entire leftover job. The optimal solution of the LP Relaxation to (\ref{migration_scheduling_allinone}) can utilize these $g^{r_i}$ idle VMs for hosting a fractional leftover job. We have $\sum_{l\in \cup_{\mathcal{J}}\widetilde{\mathcal{U}}_j}g^k \overline{I^l_{i,0}}(t) = \sum_{l\in \cup_{\mathcal{J}}\widetilde{\mathcal{U}}_j}g^k I^l_{i,0}(t)^{(a)} + g^{r_i}, \forall i\in\mathcal{D}_L$. There are $\sum_{i\in\mathcal{D}_L} g^{r_i}$ more VMs provided by data centers in $\mathcal{D}_H$ under Algorithm \ref{alg:scheduling} than the optimal solution of the LP Relaxation to (\ref{migration_scheduling_allinone}).

{\small\begin{align*}
& \widetilde{\varphi_2}^{(a)}(t) - \widetilde{\varphi_2}^*(t) \\
= & - \sum_{i\in \mathcal{D}_L}[ V\beta_i(t) - \frac{(w^{k^*})^2q_{j^*}^{k^*}(t) + Z_{j^*}^{k^*}(t)}{g^{k^*}} ]g^{r_i} \\
& + \sum_{l\in \cup_{\mathcal{J}}\widetilde{\mathcal{U}}_j}\sum_{i\in\mathcal{D}_L}[\frac{(w^{k^*})^2q_{j^*}^{k^*}(t)+ Z_{j^*}^{k^*}(t)}{g^{k^*}} \\
& + V\sum_{h\neq i} \lambda_{h}I^l_{h,0}(t-1) \cdot \phi^k ][g^k I^l_{i,0}(t)^{(a)}- g^k \overline{I^l_{i,0}}(t)]\\
& + \sum_{l\in \cup_{\mathcal{J}}\widetilde{\mathcal{U}}_j} \sum_{i\in\mathcal{D}_H} [V\beta_i(t) + V\sum_{h\neq i} \lambda_{h}I^l_{h,0}(t-1) \cdot \phi^k ]\\
& \cdot [g^k I^l_{i,0}(t)^{(a)}- g^k \overline{I^l_{i,0}}(t)] \\
= & \sum_{i\in \mathcal{D}_L}g^{r_i}[-V\beta_i(t) - V\sum_{h\neq i} \lambda_{h}I^l_{h,0}(t-1) \cdot \phi^k] \\
& + \sum_{l\in \cup_{\mathcal{J}}\widetilde{\mathcal{U}}_j} \sum_{i\in\mathcal{D}_H} [V\beta_i(t) + V\sum_{h\neq i} \lambda_{h}I^l_{h,0}(t-1) \cdot \phi^k ]\\
& \cdot [ g^k I^l_{i,0}(t)^{(a)} - g^k \overline{I^l_{i,0}}(t)] \\
\le & \sum_{i\in \mathcal{D}_L}g^{r_i}(-V\beta_{min}) + \sum_{i\in \mathcal{D}_H}\sum_{l\in \cup_{\mathcal{J}}\widetilde{\mathcal{U}}_j}[ g^k I^l_{i,0}(t)^{(a)} - g^k \overline{I^l_{i,0}}(t)]V\beta_{max} \\
& + \sum_{i\in \mathcal{D}_H}\sum_{l\in \cup_{\mathcal{J}}\widetilde{\mathcal{U}}_j}[ g^k I^l_{i,0}(t)^{(a)} - g^k \overline{I^l_{i,0}}(t)]V\sum_{h\neq i} \lambda_{h}I^l_{h,0}(t-1) \cdot \phi^k \\
& + \sum_{i\in \mathcal{D}_L}\sum_{l\in \cup_{\mathcal{J}}\widetilde{\mathcal{U}}_j}[ g^k I^l_{i,0}(t)^{(a)} - g^k \overline{I^l_{i,0}}(t)]V\sum_{h\neq i} \lambda_{h}I^l_{h,0}(t-1) \cdot \phi^k
\\
\le & \sum_{i\in \mathcal{D}_L}g^{r_i}(-V\beta_{min}) + \sum_{i\in \mathcal{D}_L}g^{r_i}V\beta_{max} \\
& + \sum_{i\in \mathcal{D}_L}g^{r_i} V \lambda_{max}\phi_{max} - \sum_{i\in \mathcal{D}_L}g^{r_i} V\lambda_{min}\phi_{min} \\
\le & VJg^{max}[\beta_{max}+ \lambda_{max}\phi_{max} - \beta_{min} -\lambda_{min}\phi_{min}].
\end{align*}
}

Summarizing the above two cases, we have
$\widetilde{\varphi_2}^{(a)}(t) \le \widetilde{\varphi_2}^*(t) + C$, where $C = VJg^{max}\cdot \max\{ \frac{\alpha^{max}}{g^{min}} - \beta_{min} ,\beta_{max}+ \lambda_{max}\phi_{max} - \beta_{min} -\lambda_{min}\phi_{min}\}$.


\section{Detail Proof of Claim 4} \label{sec:appendix_c}

\noindent {\em Proof: }
Substitute $b^l = g^{k_l}\cdot[\beta_{h_l}(t)- \lambda_{h_l}\phi^{k_l}]$ for $l \in \widetilde{\mathcal{U}}_j$, $b^l = [(w^{k_j^*})^2q_j^{k_j^*}(t) + Z_j^{k_j^*}(t)]/V$ for $l\in \mathcal{U}_j^{k_j^*,max}$, $s_i = \beta_i(t)$ into the objective function of (\ref{winner_determination}), and replace $x_i^l$'s using $I_{i,0}^l$'s according to the following correspondence: for a new job $l \in \mathcal{U}_j^{k_j^*}$, $x_i^l = I_{i,0}^l$; for a new job $l\in\mathcal{U}_j^{k_j^*,max} - \mathcal{U}_j^{k_j^*}$, $x_i^l = 0$; for a leftover job $l\in\widetilde{\mathcal{U}}_j$, $x_i^l = I_{i,0}^l, \forall i\neq h_l$; $x_{h_l}^l = 0$. We get

{\small\begin{align*}
& - \sum_{j\in\mathcal{J}}\sum_{l\in \mathcal{U}_j^{k_j^*,max}}\sum_{i=1}^J[s_i\cdot g^{k_l}x_i^l  - b^l \cdot x_i^l ] \\
& - \sum_{j\in\mathcal{J}}\sum_{l\in \widetilde{\mathcal{U}}_j}\sum_{i=1}^J [s_i\cdot g^{k_l}x_i^l - b^l \cdot x_i^l ] \\
 = & -\sum_{j\in\mathcal{J}}\sum_{l\in \mathcal{U}_j^{k_j^*}}\sum_{i=1}^J \frac{1}{V}g^{k_l}I_{i,0}^l[V\beta_i(t)  - \frac{(w^{k_j^*})^2q_j^{k_j^*}(t) + Z_j^{k_j^*}(t)}{g^{k_l}}] \\
& - \sum_{j\in\mathcal{J}}\sum_{l\in \widetilde{\mathcal{U}}_j}\sum_{i\neq h_l}g^{k_l}I_{i,0}^l[\beta_i(t) - (\beta_{i_l}(t)- \lambda_{h_l}\phi^{k_l})] \\
= & -\sum_{j\in\mathcal{J}}\sum_{l\in \mathcal{U}_j^{k_j^*}}\sum_{i=1}^J \frac{1}{V}g^{k_l}I_{i,0}^l[V\beta_i(t)  - \frac{(w^{k_j^*})^2q_j^{k_j^*}(t) + Z_j^{k_j^*}(t)}{g^{k_l}}] \\
& - \sum_{j\in\mathcal{J}}\sum_{l\in \widetilde{\mathcal{U}}_j}\sum_{i\neq h_l}g^{k_l}I_{i,0}^l[\beta_i(t) + \lambda_{h_l}\phi^{k_l}] - \sum_{j\in\mathcal{J}}\sum_{l\in \widetilde{\mathcal{U}}_j}\sum_{i= h_l}g^{k_l}I_{i,0}^l[\beta_{h_l}(t)]\\
& + \sum_{j\in\mathcal{J}}\sum_{l\in \widetilde{\mathcal{U}}_j}\sum_{i=1}^J g^{k_l}I_{i,0}^l\beta_{h_l}(t) \\
= & -\frac{1}{V}\widetilde{\varphi_2}(t) + \sum_{j\in\mathcal{J}}\sum_{l\in \widetilde{\mathcal{U}}_j}g^{k_l}\beta_{h_l}(t).
\end{align*}
}

\section{Detail Proof of Claim 5} \label{sec:appendix_d}

\noindent {\em Proof:} It is obvious that {\small$\sum_{j=1}^J\sum_{l\in \widetilde{\mathcal{U}}_j\cup\mathcal{U}_j^{k_j^*,max}}\sum_{i=1}^Jx_i^{l}(z) \cdot \nu_i^l = \sum_{j=1}^J\sum_{l\in \widetilde{\mathcal{U}}_j\cup\mathcal{U}_j^{k_j^*,max}}\sum_{i=1}^J \widehat{x}_i^{l} \cdot \nu_i^{l+}$}.  We have {\small$\sum_{j=1}^J\sum_{l\in \widetilde{\mathcal{U}}_j\cup\mathcal{U}_j^{k_j^*,max}}\sum_{i=1}^J \widehat{x}_i^{l} \cdot \nu_i^{l+} \ge (1-\delta) \max_{\mathbf{x}\in \mathcal{P}}\sum_{j=1}^J\sum_{l\in \widetilde{\mathcal{U}}_j\cup\mathcal{U}_j^{k_j^*,max}}\sum_{i=1}^J x_i^{l}\cdot \nu_i^{l+}$}
since our approximation algorithm verifies the integrality gap between WDP and LPR, here $\mathcal{P}$ is the feasible region of the LPR. {\small$(1-\delta) \max_{\mathbf{x}\in \mathcal{P}}\sum_{j=1}^J\sum_{l\in \widetilde{\mathcal{U}}_j\cup\mathcal{U}_j^{k_j^*,max}}\sum_{i=1}^J x_i^{l}\cdot \nu_i^{l+} \ge (1-\delta) \max_{\mathbf{x}\in \mathcal{P}}\sum_{j=1}^J\sum_{l\in \widetilde{\mathcal{U}}_j\cup\mathcal{U}_j^{k_j^*,max}}\sum_{i=1}^J x_i^{l}\cdot \nu_i^{l}$} as $\nu_i^{l+} \ge \nu_i^l$. Combining the above three relation, we can get the inequality in Claim 5.

\end{document}